\documentstyle[prd,aps,floats,tighten,epsfig,graphicx,color]{revtex}

\newcommand{\beq}{\begin{equation}}
\newcommand{\eeq}{\end{equation}}
\newcommand{\bea}{\begin{eqnarray}}
\newcommand{\ena}{\end{eqnarray}}

\newcommand{\ie}{{\it i.e.}}
\newcommand{\lsim}{\mathrel{\mathop{\kern 0pt \rlap
{\raise.2ex\hbox{$<$}}}
\lower.9ex\hbox{\kern-.190em $\sim$}}}
\newcommand{\gsim}{\mathrel{\mathop{\kern 0pt \rlap
{\raise.2ex\hbox{$>$}}}
\lower.9ex\hbox{\kern-.190em $\sim$}}}

\newcommand{\hepph}[1]{{\tt hep-ph/#1}}
\newcommand{\astroph}[1]{{\tt astro-ph/#1}}
\newcommand{\prep}[3]{Phys.\ Rep.\ {\bf #1}, #3 (#2)}
\newcommand{\plb}[3]{Phys.\ Lett.\ B\ {\bf #1}, #3 (#2)}

\renewcommand{\apj}[3]{Astrophys.\ J.\ {\bf #1}, #3 (#2)}
\newcommand{\aeta}[3]{Astron.\ {\&}\ Astrophys.\ {\bf #1}, #3 (#2)}
\newcommand{\pr}[3]{Phys.\ Rev.\ {\bf #1}, #3 (#2)}
\renewcommand{\prl}[3]{Phys.\ Rev.\ Lett. {\bf #1}, #3 (#2)}
\renewcommand{\prd}[3]{Phys.\ Rev.\ D\ {\bf #1}, #3 (#2)}

\newcommand{\href}[2]{#1}

\newcommand{\rb}{\mbox{$\rho_{\rm \, b}$}}
\newcommand{\pb}{\mbox{$P_{\rm \, b}$}}
\newcommand{\rp}{\mbox{$\rho_{\phi}$}}
\newcommand{\rpp}{\mbox{$\rho'_{\phi}$}}
\newcommand{\rps}{\mbox{$\rho''_{\phi}$}}
\newcommand{\pp}{\mbox{$P_{\phi}$}}

\definecolor{cyan}{cmyk}{1.,0.,0.,0.5}
\definecolor{magenta}{cmyk}{0.,1.,0.,0.5}
\definecolor{verdatre}{cmyk}{0.5,0.,0.5,0.5}
\definecolor{yellow}{cmyk}{0.,0.,0.2,0.0}
\definecolor{rouge}{cmyk}{0.,0.4,0.6,0.0}
\definecolor{orange}{cmyk}{0.,0.5,0.5,0.}
\definecolor{violet}{rgb}{0.5,0.,0.5}

\begin{document}

\title{Galactic Halos of Fluid Dark Matter
\footnote{CERN-TH/2003-015, LAPTH--961/03}}
\vskip 1.cm
\author{
Alexandre Arbey$^{\rm a,b}$
\footnote{E--mail: arbey@lapp.in2p3.fr, lesgourg@lapp.in2p3.fr,
salati@lapp.in2p3.fr},
Julien Lesgourgues$^{\rm c,a}$ and Pierre Salati$^{\rm a,b}$
}
\vskip 0.5cm
\address{
\begin{flushleft}
a) Laboratoire de Physique Th\'eorique LAPTH, B.P.~110, F-74941
Annecy-le-Vieux Cedex, France.\\
b) Universit\'e de Savoie, B.P.~1104, F-73011 Chamb\'ery Cedex,
France.\\
c) Theoretical Physics Division, CERN, CH-1211 Gen\`eve 23, 
Switzerland. 
\end{flushleft}
}
\maketitle
\vskip 0.5cm
\begin{abstract}
Dwarf spiral galaxies -- and in particular the prototypical
DDO 154 -- are known to be completely dominated by an unseen
component. The putative neutralinos -- so far the favored
explanation for the astronomical dark matter -- fail to reproduce
the well measured rotation curves of those systems because these
species tend to form a central cusp whose presence is not supported
by observation.
We have considered here a self--coupled charged scalar field as
an alternative to neutralinos and investigated whether a Bose
condensate of that field could account for the dark matter inside
DDO 154 and more generally inside dwarf spirals. The size of the
condensate turns out to be precisely determined by the scalar mass
$m$ and self--coupling $\lambda$ of the field. We find actually that
for $m^{4} / \lambda \sim$ 50 -- 75 eV$^{4}$, the agreement with the
measurements of the circular speed of DDO 154 is impressive whereas
it lessens for larger systems.
The cosmological behavior of the field is also found to be
consistent -- yet marginally -- with the limits set by BBN on
the effective number of neutrino families.
We conclude that classical configurations of a scalar and
self--coupled field provide a possible solution to the astronomical
dark matter problem and we suggest further directions of research.
\end{abstract}
%
\vskip 1.cm

\section{Introduction}
\label{sec:introduction}

\noindent
After many years of global consensus on the fact that dark matter
consists in Weakly Interacting Massive Particles (WIMPs) -- like, for
instance, the lightest neutralino in the Minimal Supersymmetric
Standard Model (MSSM) -- there is still no strong evidence in favor of
WIMPS, neither from bolometer experiments designed for direct
detection, nor from the observation of cosmic rays, a fraction of
which could consist in WIMPs annihilation products. This absence of
experimental constraints on dark matter from the particle physics side 
leaves the door wide open for alternative descriptions of the hidden
mass of the Universe.

\vskip 0.1cm
\noindent
Moreover, in the past three years, there has been a lot of controversy 
concerning the small--scale inhomogeneities of the WIMPs density.
Indeed, many recent N--body simulations of structure formation in
the Universe suggested that any 
dark matter component modelized as a gas of free particles
-- such as WIMPs -- tends to cluster excessively on scales of order
1~kpc and smaller. This would result in cuspy density profiles at 
galactic centers, while most rotation curves indicate a smooth core 
density \cite{moore}. 
Many galaxies even seem to be dominated by baryons near their 
center, with a significant dark matter fraction only at large radii. 
This clearly contradicts the results from current N--body simulations,
in which the dark matter density is strongly enhanced at the center 
of the halo with respect to its outskirts.

\vskip 0.1cm
\noindent
This argument was attacked by Weinberg and Katz \cite{weinberg}, 
who stressed the importance of
including the baryon component in N--body simulations. Indeed, the
baryon dissipative effects could be responsible for a smoothing 
of the central dark matter cusp in the early Universe. This possible
solution to the dark matter crisis was discarded later by 
Sellwood \cite{sellwood},
who found opposite results in his simulation.

\vskip 0.1cm
\noindent
Apart from the central cusp problem, N--body simulations
raised some secondary issues \cite{moore}. First, a clumpy halo could
generate some tidal effects that could break the spatial coherence of
the disk.  Second, the predicted number of satellite galaxies around
each galactic halo is far beyond what we see around the Milky Way.
Third, the dynamical friction between dark matter particles and
baryons should freeze out the spinning motion of baryonic bars in
barred galaxies.  All these arguments are still unclear, because they
seem to depend on the resolution under which simulations are carried
\cite{font,valenzuela}, and also because of our ignorance of what
could be the light--to--mass ratio inside small dark matter clumps. In
addition, the predicted number of satellite galaxies doesn't seem to
be in contradiction with constraints from microlensing
\cite{kochanek}.

\vskip 0.1cm
\noindent
Should these various problems be confirmed or not,
it sounds reasonable to explore alternatives to the WIMPs model, or
more generally speaking, to any description based on a gas of free
particles. This can be done in various ways: for instance, one can
introduce some deviations from a perfect thermal phase--space
distribution \cite{brandenberger}, or add a self--coupling between
dark matter particles \cite{sicdm}. A more radical 
possibility is to drop the assumption that dark matter is governed
by the laws of statistical thermodynamics. This would be the case
if dark matter consisted in a classical scalar field, coherent on very
large scales, and governed by the Klein--Gordon equation of motion.

\vskip 0.1cm
\noindent
This framework should be clearly distinguished from other models of bosonic
dark matter, like those based on heavy bosons -- for instance, sneutrinos
-- or axions. In the first case, the Compton wavelength $\hbar/(mc)$
of an individual particle is much smaller than the typical interparticle
distance while in the second case, for axion masses of order $10^{-6}$~eV,
it is still much smaller than the typical size of a galaxy. So, in these 
examples, the bosons can be described on astronomical scales like
a gas of free particles in statistical equilibrium. It follows that the halo 
structure cannot be distinguished from that of standard WIMPs.

\vskip 0.1cm
\noindent
A coherent scalar field configuration governed by the Klein--Gordon
and Einstein equations is nothing but a self--gravitating Bose
condensate. Such condensates span over scales comparable to the
De Broglie wavelength, $L = \hbar / p$. In the case of free bosons
-- \ie, with a quadratic scalar potential -- the momentum $p$ is
of order $m v_{\rm esc}$
where $v_{\rm esc}$ is the escape velocity from the system.
Typical examples are boson stars \cite{ruffini,lee,jetzer,liddle}, 
for which the characteristic orders of magnitude discussed in the
literature  are, for instance, $m \sim 10$~GeV
and $v_{\rm esc} \sim c$, leading to a radius as tiny as
$L \sim 10^{-14}$~cm.
Even for axions, which have a much smaller
mass and an escape velocity given by the motion of stars in a galaxy
-- $v_{\rm esc} \sim 100$~km/s -- the De Broglie wavelength is only of 
order $L = 100$~km, so that on galactic scales, the medium can be
treated as a gas.
In order to obtain a galactic halo described by the Klein--Gordon
equation, one should consider masses of order $m = \hbar / (L v_{\rm
esc})$ where $v_{\rm esc} \sim 100$~km/s and $L \sim 10$~kpc. This
yields $m \sim 10^{-23}$~eV. Such an ultra--light scalar field was
called ``fuzzy dark matter'' by Hu \cite{hu}, who discussed its
overall cosmological behavior. In some previous works, we focused on a
variant of this model in which the ultra--light scalar field is
complex -- then, the conserved number associated with the $U(1)$
global symmetry helps in stabilizing the condensate against
fragmentation \cite{lee,liddle}. In \cite{ajp1} -- thereafter Paper~I
-- we compared the rotation curves predicted by this model with some
data from spiral galaxies. In \cite{ajp2} -- thereafter Paper~II -- we
simulated the cosmological evolution of the homogeneous background of
such a field. The model seems to be quite successful in explaining the
rotation curves, but it has two caveats. First, such a low mass is
very difficult to implement in realistic particle physics models.
The second problem is related to the fact that because of the $U(1)$
symmetry, the field carries a conserved quantum number. As explained
in Paper~II, the value derived from cosmological considerations for
the density of this quantum number does not seem consistent with that
inferred from astrophysical arguments.

\vskip 0.1cm
\noindent
These caveats motivate the introduction of a quartic self--coupling term
in the scalar potential. In that case, it is already known from boson
stars that for the same value of the mass, the self--coupling constrains
the field to condensate on much larger scales \cite{colpi,liddle} --
the size of the self--gravitating configurations is still given by
$L = \hbar / p$, but in presence of a self--coupling, the momentum
cannot be identified with $m v_{\rm esc}$. Then, without recurring to
ultra--light masses, we may still describe the galactic halos with a Bose
condensate. A massive scalar field with quartic -- or close to quartic --
self--coupling was proposed as a possible dark matter candidate by Peebles,
who called it ``fluid dark matter'' \cite{peebles}.
In this paper, we will study a variant of fluid dark matter in which
the quartically self--coupled massive scalar field is complex, still
for stability reasons\footnote{In contrast, a scalar field dark matter
model in which the field is real and unstable is discussed in
\cite{tkachev}.}. Some cosmological properties of such a field were already discussed in \cite{boyle}.

\vskip 0.1cm
\noindent
We will focus mainly on galaxy rotation curves, assuming that the
dark matter halos are the self--gravitating, aspherical and stable
equilibrium configurations of our scalar field in the presence
of a baryonic matter distribution -- stellar disk, HI gas, etc...
We will present here the first solution of this problem. However,
we should stress that some different models in which the rotation
curves are also seeded by a coherent scalar field were studied
previously by
Schunk \cite{schunk} -- with a vanishing scalar potential,
Goodman \cite{goodman} -- with a repulsive self-interaction,
Matos et al. \cite{mexicans},
Nucamendi et al. \cite{nucamendi}, 
Wetterich \cite{wetterich},
Urena-Lopez and
Liddle \cite{ULL}.
In some of these papers, and also in many other recent proposals
-- see for instance \cite{coincidence} -- the main goal is to try to
solve simultaneously the dark energy and dark matter problems, assuming
that a quintessence field can cluster on galactic scales. This raises
some subtle issues, like the existence of a scale--dependent equation
of state. At the present stage, we do not have such an ambition, and we 
will focus only on the dark matter problem.

\vskip 0.1cm
\noindent
In section~\ref{sec:basic_equations},
we write the Einstein and the Klein--Gordon
equations which govern the scalar field and the gravitational potential
distributions in the presence of a given baryonic matter density.  We
will see that these equations can be combined into a single
non--linear Poisson equation.
The solutions are technically difficult to find, first, due to the
non--linearity, and second, because some boundary conditions are given
at the center, some others at infinity. So, it is not possible to
follow a lattice approach, in which one would start from a particular
point and integrate numerically grid point by grid point.  However
we present in section~\ref{sec:method} a recursive method which
allows to find all the exact solutions after a few iterations.
In section~\ref{sec:rotation_curves},
we compare the galaxy rotation curves obtained
in this way with some observational data. We lay a particular
emphasis on the dwarf spiral galaxy DDO~154, for which the rotation
curve is among the most difficult to explain with usual dark matter
profiles. We will see that a mass--over--self--coupling ratio
$m^{4} / \lambda \simeq 50$ $({\rm eV})^4$ provides a very good fit
to the DDO~154 rotation curve -- however at the expense of poor
fits to the largest spiral galaxies.
Because the scalar field condensates inside the gravitational
potential wells of baryons and strengthens them, the question of
its effects on the inner dynamics of the solar system naturally
arises. We derive in section~\ref{sec:solar_system} the modification
of the solar attraction in the presence of the self--interacting
scalar field under scrutiny and show that an anomalous acceleration
appears that is constant and that points towards the Sun. We
investigate the limit set on our model by the Pioneer radio data.
In Paper~II, we studied the cosmological behavior of a homogeneous
scalar field that was assumed to play the role of dark matter at
least from the time of matter--radiation equality until today. This
analysis is updated in section~\ref{sec:cosmology} where we
specifically assume $m^{4} / \lambda \simeq 50$ $({\rm eV})^4$.
Such a large value point towards a large total density of the Universe
during radiation domination which is at the edge of the current
bounds set in particular by BBN on cosmological parameters.
%
%
The last section is devoted to a discussion of the strong and weak aspects
of our alternative dark matter model. We finally suggest some further
directions of investigation beyond the simple but restrictive framework
of isolated bosonic configurations.

\section{Gravitational behavior}
\label{sec:basic_equations}

\noindent
The complex scalar field $\phi$ under scrutiny in this article
is associated to the Lagrangian density
\beq
{\cal L} \; = \;
g^{\mu \nu} \, \partial_{\mu} \phi^{\dagger} \, \partial_{\nu} \phi
\; - \; V \left( \phi \right) \;\; ,
\eeq
where the $U(1)$ invariant potential $V$ includes both quadratic
and quartic contributions
\beq
V \left( \phi \right) \; = \; m^{2} \, \varphi^{\dagger} \phi
\; + \; \lambda \, \left\{ \phi^{\dagger} \phi \right\}^{2} \;\; .
\eeq
The gravitational behavior of the system follows the standard GR
equations whilst the field $\phi$ satisfies the Klein--Gordon
equation
\beq
{\displaystyle \frac{1}{\sqrt{- g}}} \; \partial_{\mu} \left\{
\sqrt{- g} \, g^{\mu \nu} \, \partial_{\nu} \phi \right\}
\; + \; {\displaystyle \frac{\partial V}{\partial \phi^{\dagger}}}
\; = \; 0 \;\; ,
\label{KG1}
\eeq
where $g_{\mu \nu}$ denotes the metric.
%
We would like to investigate to which extent the scalar field $\phi$
may account for the dark matter inside galaxies. The problem simplifies
insofar as the gravitational fields at stake are weak and static. In this
quasi--Newtonian limit of general relativity, deviations from the Minkowski
metric $\eta_{\mu \nu} = diag \left\{1,-1,-1,-1 \right\}$ are accounted
for by the perturbation tensor $h_{\mu \nu}$ -- from now on, we use
the convention $c=1$. The Newtonian gravitational
potential $\Phi = h_{00} / 2$ is actually a small quantity of order
$v_{\rm esc}^{2} \sim 10^{-7} - 10^{-6}$, where $v_{\rm esc}$ denotes
the escape velocity. Our analysis is based on an expansion up to first
order in $\Phi$.
%
The baryonic content of galaxies is described through the
energy--momentum tensor
\beq
T^{\mu \nu} \; = \; \left( \rb + \pb \right) U^{\mu} U^{\nu} \; - \;
\pb \, g_{\mu \nu} \;\; ,
\eeq
where $U^{\mu} = \{ 1 , \vec{v} \}$. Baryons behave as dust
with non--relativistic velocities. Actually, because galaxies are
virialized systems -- hence the assumption of static gravitational
fields -- the spatial velocity $v$ is a small quantity of order
$v_{\rm esc} \sim \sqrt{\Phi}$. The kinetic pressure--to--mass
density ratio is even more negligible since
$\pb / \rb \sim v^{2} \sim v_{\rm esc}^{2} \sim \Phi \ll 1$.
%
We are
interested in classical configurations where the field $\phi$
is in a coherent state such as
\beq
\phi \left\{ \vec{x} , t \right\} \; = \;
{\displaystyle \frac{\sigma \left( \vec{x} \right)}{\sqrt{2}}} \,
\exp \left( - \, i \, \omega \, t \right) \;\; .
\label{configuration_scalar}
\eeq
Indeed, one can prove that
all stable spherically symmetric configurations can be parameterized 
in that way \cite{lee}.
The time--derivative $\partial_{0} \phi$ equals $- i \, \omega \, \phi$,
whereas the space--derivative $\partial_{i} \phi$ is of order 
$\phi / L$
where $L$ is the physical length of the configuration. That length
-- which is related to the parameters $m$ and $\lambda$ of the potential
-- is required to be $\sim$ 1 -- 100 kpc to account for the galactic
dark matter. On the other hand, we shall see later that 
$\omega$ is very close to the mass $m$,
which is numerically found to be in the ballpark of a fraction of eV.
We readily infer a ratio
\beq
{\displaystyle \frac{\partial_{i} \phi}{\partial_{0} \phi}}
\sim {\displaystyle \frac{1}{m \, L}} \; = \;
6.4 \times 10^{-27} \;
\left\{ {\displaystyle \frac{1 \, {\rm eV}}{m}} \right\} \;
\left\{ {\displaystyle \frac{1 \, {\rm kpc}}{L}} \right\} \;\; .
\eeq
So, the space derivative $\partial_{i} \phi$ of the scalar
field can be safely neglected throughout the analysis. 

\vskip 0.1cm
\noindent
In its weak--field limit, general relativity becomes a gauge theory.
By conveniently choosing the gauge of harmonic coordinates in which
the metric perturbation $h_{\mu \nu}$ satisfies the condition
\beq
\partial_{\mu} h^{\mu}_{\;\; \alpha} \; = \; \frac{1}{2} \,
\partial_{\alpha} \left\{ h^{\mu}_{\;\; \mu} \right\} \;\; ,
\label{harmonic}
\eeq
the GR equations simplify into
\beq
\Box h_{\mu \nu} \; = \; - \, 16 \, \pi \, G \, S_{\mu \nu} \;\; .
\label{dalembert}
\eeq
The effective source $S_{\mu \nu}$ is related to the energy--momentum
tensor $T_{\mu \nu}$ through
\beq
S_{\mu \nu} \; = \; T_{\mu \nu} \, - \,
\frac{1}{2} \, g_{\mu \nu} \, T^{\lambda}_{\;\; \lambda} \;\; .
\eeq
%
In the propagation eq.~(\ref{dalembert}), the source $S_{\mu \nu}$
is computed in flat space while the metric perturbation $h_{\mu \nu}$ is
of order $\Phi$. If the dark matter inside galaxies is understood as some
classical configuration of the field $\phi$, $S_{\mu \nu}$ should take
into account both baryonic population -- stars and gas -- and scalar
condensate. In the Newtonian limit where gravito--magnetic effects
are disregarded, the non--relativistic velocities of baryons can be
neglected. The only non--vanishing components of the baryonic source
tensor are
\beq
S^{\rm \; b}_{00} \; = \; {\displaystyle \frac{\rb}{2}}
{\hskip 1.cm}{\rm and}{\hskip 1.cm}
S^{\rm \; b}_{ij} \; = \; - \, \eta_{ij} \,
{\displaystyle \frac{\rb}{2}} \;\; .
\eeq
%
Assuming that eq.~(\ref{configuration_scalar}) describes the scalar
field configuration and disregarding the space--derivatives
$\partial_{i} \phi$ leads to the source components
\beq
S^{\; \varphi}_{00} \; = \; \omega^{2} \sigma^{2} \, - \, V
{\hskip 1.cm}{\rm whilst}{\hskip 1.cm}
S^{\; \varphi}_{ij} \; = \; - \, \eta_{ij} \, V \;\; ,
\eeq
where the potential is
\beq
V \left( \sigma \right) \; = \;
{\displaystyle \frac{m^{2}}{2}}   \, \sigma^{2} \; + \;
{\displaystyle \frac{\lambda}{4}} \, \sigma^{4} \;\; .
\eeq
%
The well--known solution of the Lienard and Wiechert retarded
potentials satisfies the propagation eq.~(\ref{dalembert}).
We readily conclude that the metric does not contain any
space--time component $h_{0i}$ and may be expressed at this stage
as
\beq
d\tau^{2} \; = \; \left( 1 \, + \, 2 \Phi \right) dt^{2} \; - \;
\left( 1 \, - \, 2 \Psi \right) \delta_{ij} \, dx^{i} dx^{j} \;\; ,
\eeq
where the static potentials $\Phi$ and $\Psi$ are given by
integrals over the source distribution ${\cal D}$ of the baryonic
and scalar mass densities
\beq
\Phi (\vec{x}) \; = \; - \, G \, {\displaystyle \int_{\cal D}} \,
{\displaystyle \frac{d^3 \vec{y}}{|\vec{x} - \vec{y}|}} \,
\left\{ \rb(\vec{y}) + \rpp(\vec{y}) \right\} \;\; ,
\eeq
and 
\beq
\Psi (\vec{x}) \; = \; - \, G \, {\displaystyle \int_{\cal D}} \,
{\displaystyle \frac{d^3 \vec{y}}{|\vec{x} - \vec{y}|}} \,
\left\{ \rb(\vec{y}) + \rps(\vec{y}) \right\} \;\; .
\eeq
The densities $\rpp$ and $\rps$
are respectively defined by
\beq
\rpp \; = \; 2 \, \omega^{2} \sigma^{2} \, - \, m^{2} \sigma^{2}
\, - \, {\displaystyle \frac{\lambda}{2}} \, \sigma^{4} \;\; ,
\eeq
and 
\beq
\rps \; = \; m^{2} \sigma^{2} \, + \,
{\displaystyle \frac{\lambda}{2}} \, \sigma^{4} \;\; .
\eeq

\vskip 0.1cm
\noindent
The potentials $\Phi$ and $\Psi$ are different {\it a priori}. A careful
inspection of the Klein--Gordon equation will eventually show that
they are actually equal. The latter may be written as
\beq
\left( 1 \, - \, 2 \Phi \right) \, \ddot{\phi} \; - \;
\left( 1 \, + \, \Phi \, - \, 3 \Psi \right)^{-1} \, \partial_{i}
\left\{ \left( 1 \, + \, \Phi \, - \, \Psi \right) \,
\partial_{i} \phi \right\} \; + \;
{\displaystyle \frac{\partial V}{\partial \phi^{\dagger}}}
\; = \; 0 \;\; .
\label{KG2}
\eeq
The space--dependent term is some 53 orders of magnitude smaller
than its time--dependent counterpart and we can safely disregard
it so that relation~(\ref{KG2}) simplifies into
\beq
\lambda \sigma^{2} \; = \;
\left( 1 \, - \, 2 \Phi \right) \omega^{2} \, - \, m^{2} \;\; ,
\eeq
where the configuration~(\ref{configuration_scalar}) has been assumed.
The scalar field is in a classical state
that may be pictured as a Bose condensate on the boundaries of which
the gravitational potential is
\beq
\Phi_{0} \; = \; \frac{1}{2}
\left( 1 \, - \, {\displaystyle \frac{m^{2}}{\omega^{2}}} \right) \;\; .
\eeq
Because the potential $\Phi_{0}$ is a small quantity, the pulsation
$\omega$ is very close to the mass $m$. The scalar field
essentially vanishes outside the condensate whereas its inner value
is directly related to the gravitational potential $\Phi$ through
\beq
\lambda \sigma^{2} \; = \; 2 \,
\left( \Phi_{0} - \Phi \right) \, \omega^{2} \; \simeq \; 2 \,
\left( \Phi_{0} - \Phi \right) \, m^{2} \;\; .
\eeq
%
This relation has important consequences. To commence, the densities
$\rpp = m^{2} \sigma^{2} \, \left\{ 1 \, + \, 3 \Phi_{0} \, - \, \Phi \right\}$
and
$\rps = m^{2} \sigma^{2} \, \left\{ 1 \, + \,   \Phi_{0} \, - \, \Phi
\, - \, 2 \Psi \right\}$
become both equal to $\rp \simeq m^{2} \sigma^{2}$ at lowest order in the
potentials. Then $\Phi \equiv \Psi$ and the metric simplifies. It is
straightforward to show that it readily satisfies the gauge
condition~(\ref{harmonic}). The scalar field density may be expressed
as a difference between the gravitational potentials inside and on the
boundary of the scalar field condensate
\beq
\rp \; = \;
{\displaystyle \frac{2 \, m^{4}}{\lambda}} \,
\left( \Phi_{0} - \Phi \right) \,
{\cal H} \left( \Phi_{0} - \Phi \right) \;\; ,
\eeq
where ${\cal H}(x) = 1$ for $x>0$ and ${\cal H}(x) = 0$ elsewhere.
This leads to the Poisson equation
\beq
\Delta \Phi \, = \; 4 \, \pi \, G \, \rb \;\;
+ \,
8 \, \pi \, G \, {\displaystyle  \frac{m^{4}}{\lambda}} \,
\left( \Phi_{0} - \Phi \right) \, {\cal H} \left( \Phi_{0} - \Phi \right)
\;\; .
\label{poisson1}
\eeq
Inside the condensate, gravity turns out to be effectively modified
by the presence of the scalar density $\rp$ whereas the conventional
Poisson equation is recovered outside. Defining the Planck mass through
$M_{\rm P} = 1 / \sqrt{G}$, we derive a typical scale of
\beq
L^{2} \; = \; {\displaystyle \frac{\lambda}{8 \, \pi}} \,
{\displaystyle \frac{M_{\rm P}^{2}}{m^{4}}} \; = \;
{\displaystyle \frac{\Lambda}{m^{2}}} \;\; ,
\label{scale1}
\eeq
for the scalar field configurations in which we are interested.
The dimensionless constant $\Lambda$ has been introduced by \cite{colpi}
in their analysis of self--interacting boson stars.
The scale $L$ is related to the mass $m$ and the quartic coupling
$\lambda$ through
\beq
L \; \simeq \; 1.6 \; {\rm kpc} \;
\left\{ {\displaystyle \frac{\lambda}{10^{-2}}} \right\}^{1/2} \;
\left\{ {\displaystyle \frac{1 \, {\rm eV}}{m}} \right\}^{2} \;\; ,
\label{scale2}
\eeq
so that values of the mass in the ballpark of the eV may well
be compatible with a size $L$ of order a few kiloparsecs. As
already noticed by \cite{colpi}, the space--dependent term in
the Klein--Gordon eq.~(\ref{KG2}) is actually suppressed by
a factor of $\Lambda$ which, in our case, reaches values as large
as $\sim 10^{53}$. The key feature of the scalar field configurations
at stake is the existence of a unique scale $L$ that depends only
on the parameters $m$ and $\lambda$ of the potential $V$. 

\vskip 0.1cm
\noindent
A pure scalar field configuration may also be seen as a mere fluid with
mass density $\rp$. The pressure $\pp$ may be derived from the
space--space component $T_{ij} = - \, \eta_{ij} \, \pp$ of its
energy--momentum tensor. This leads to
\beq
\pp \; = \; {\cal L} \; = \;
g^{00} \, \dot{\phi}^{\dagger} \dot{\phi} \, - \, V \; \simeq \;
{\displaystyle \frac{m^{4}}{\lambda}} \,
\left( \Phi_{0} - \Phi \right)^{2}
\eeq
inside the condensate where $\Phi \le \Phi_{0}$. The corresponding
equation of state boils down to
\beq
\pp \; = \; {\displaystyle \frac{\lambda}{4 \, m^{4}}} \; \rp^{2}
\;\; ,
\eeq
and features the generic polytropic form $P \, = \, K \, \rho^{\Gamma}$
where $\Gamma = 1 + 1/n$ and $K = {\lambda}/{4 \, m^{4}}$ are
constants. Spherical symmetric solutions of configurations in
hydrostatic equilibrium are searched of the form
$\rho / \rho_{C} = \Theta^{n}(z)$ and
$P / P_{C} = \Theta^{n+1}(z)$ where $z = r / L$ is the dimensionless
radius. The typical scale of the polytrope depends on the central
density $\rho_{C}$ and pressure $P_{C}$ through
\beq
L^{2} \; = \; {\displaystyle
\frac{(n+1) P_{C}}{4 \, \pi \, G \, \rho_{C}^{2}}} \;\; ,
\eeq
whereas the generic function $\Theta$ satisfies the Lane--Emden
equation
\beq
{\displaystyle \frac{1}{z^{2}}} \, {\displaystyle \frac{d}{dz}}
\left\{ z^{2} \, {\displaystyle \frac{d \Theta}{dz}} \right\}
\; = \; - \, \Theta^{n} \;\; ,
\eeq
with the initial conditions $\Theta(0) = 1$ and $\Theta'(0) = 0$.
In the scalar field case, the polytropic index is $n = 1$ and
the solution
\beq
{\displaystyle \frac{\rp}{\rp_{C}}} \; = \; \Theta(z) \; = \;
{\displaystyle \frac{\sin z}{z}}
\eeq
readily obtains. It describes a spherical symmetric configuration
where the scalar field alone is bound by its own gravity. The
radius of the pure scalar field condensate is then 
$R = \pi \, L$ where the scale $L$ 
has already been derived in relations~(\ref{scale1}) and
(\ref{scale2}):
\beq
R \; = \; \pi \, L
\; = \; \pi
\left\{ {\displaystyle \frac{K}{2 \, \pi \, G}} \right\}^{1/2}
\; \equiv \; \pi
\left\{ {\displaystyle \frac{\lambda}{8 \, \pi \, G}} \right\}^{1/2}
\, {\displaystyle \frac{1}{m^{2}}}
\label{scale3}
\eeq
The effect of an aspherical distribution of baryons
on the scalar field condensate will be examined in the next section.

\section{Resolution method}
\label{sec:method}

\noindent
We would like to compute the gravitational potential $\Phi$ 
associated with any density of baryons $\rho_b$ in the galaxy, in the
presence of a scalar field condensate. 
So, we need to solve eq.~(\ref{poisson1}).
The Heaviside function renders this equation strongly non--linear:
different solutions have different surfaces where
\beq
\Phi(\vec{x})=\Phi_0 \, ,
\eeq
so the sum of two solutions is not a solution.
Nevertheless, it is possible to solve the equation with a recursive 
method. The idea is to start from an approximate solution
$\Phi^{(0)}$, and to find $\Phi^{(n)}$ from the iterations
\beq
\label{eq:iterative}
\Delta \Phi^{(n+1)} = 4 \pi G \rho_b + 8 \pi G\frac{m^4}{\lambda} 
\left(\Phi_0-\Phi^{(n)}\right) \, \, {\cal H} \! \left(\Phi_0 \! - \! 
\Phi^{(n)}\right) \;\; .
\label{poisson2}
\eeq
If, for a judicious choice of $\Phi^{(0)}$,
the $\Phi^{(n)}$'s converge towards a limit
$\Phi^{(\infty)}$, 
then the latter will be an exact solution of 
(\ref{poisson1}).

\vskip 0.1cm
\noindent
We will always work in the approximation in which the
baryonic density $\rho_b(\vec{x})$ is
axially symmetric, continuous and vanishing at infinity.
So, the induced gravitational potential should be
\beq
\left\{
\begin{array}{l}
{\rm axially~symmetric,}\\
{\rm continuous~and~twice~derivable,}\\
{\rm vanishing~at~infinity.}
\end{array}
\right.
\label{properties}
\eeq
We introduce a spherical coordinate system $(r,\theta,\varphi)$ where 
$(\theta=0)$ defines the symmetry axis. So, there will be no 
$\varphi$--dependence in the solutions, and a good way to 
find them is to perform a Legendre transformation.
Let's first illustrate this for the general Poisson equation:
\beq
\label{eq:simple}
\Delta \Phi(r,\theta) = S(r,\theta) \, ,
\eeq
\noindent 
where $S$ and $\Phi$ possess the properties of symmetry and
continuity listed previously.
If one decomposes the potential $\Phi$ and the 
source term $S$ into Legendre polynomials:
\begin{eqnarray}
\Phi(r,\theta)=\sum_{l=0}^{+\infty} P_l(\cos \theta) \Phi_l(r) \, ,
\label{recomposition}
\\
S(r,\theta)= \sum_{l=0}^{+\infty} P_l(\cos \theta) S_l(r) \, ,
\end{eqnarray}
\noindent then the $S_l$'s are found from
\beq
S_l(r)=\frac{2 l+1}{2}\int_{-1}^{+1} S(r,\theta) P_l(\cos \theta) 
d(\cos 
\theta) \; ,
\label{decomposition}
\eeq
\noindent while the $\Phi_l$'s are the solutions of the linear set of
equations
\beq
\frac{1}{r^2}\frac{d}{dr}\left( r^2 \frac{d\Phi_l}{dr}\right) - 
\frac{l(l+1)}{r^2} \Phi_l = S_l \, .
\label{diff_eq_l}
\eeq
\noindent 
The boundary conditions are given by the properties (\ref{properties})
for all $l$'s, 
\beq
\frac{d}{dr}\Phi_l(0)=0 \qquad {\rm and} \qquad 
\lim_{r \to +\infty}\Phi_l(r)=0 \, .
\eeq
So, in order to find the solution of  eq.~(\ref{diff_eq_l}), one can first
compute some Green functions $G_l$ that are continuous, null at 
infinity, with zero derivative at the center, and verifying
\beq
\frac{1}{r^2}\frac{d}{dr}\left( r^2 
\frac{\partial}{\partial r} G_l(r,u) \right) - 
\frac{l(l+1)}{r^2} G_l(r,u) = \delta(r-u)
\eeq
\noindent where $\delta$ is the Dirac function. The unique answer is
\beq
G_l(r,u)=-\frac{u}{2 l+1} \left\{ \left(\frac{r}{u}\right)^l 
{\cal H}(u-r) + \left( \frac{r}{u}\right)^{-(l+1)} {\cal H}(r-u)\right\} 
\, .
\eeq
Since $\displaystyle \Phi_l(r)=\int_0^{+\infty} S_l(u) G_l(u,r) du$, one 
finally finds
\beq
\Phi_l(r)=-\frac{r^{-(l+1)}}{2l+1}\int^r_0 S_l(u)u^{l+2}du - 
\frac{r^l}{2l+1}\int^{+\infty}_r S_l(u) u^{1-l}du \, .
\label{green_solution}
\eeq
\noindent We can still use this Green function technique in our
recursive method. Indeed, if $\Phi^{(n)}$ shares the properties 
(\ref{properties}), then we can identify the right-hand side of 
eq.~(\ref{eq:iterative}) with $S(r, \theta)$, and find 
$\Phi^{(n+1)}$ using the method described above. Then 
$\Phi^{(n+1)}$ also shares the properties 
(\ref{properties}).

\vskip 0.1cm
\noindent So, at each recursion step, we need to expand the right
hand-side of eq.~(\ref{eq:iterative}) in Legendre coefficients. 
Note that the $\Phi^{(n)}_l$'s are known from the previous iteration,
while the number $\Phi_0$ has to be imposed in some arbitrary way. In 
fact, looking again at eq.~(\ref{poisson1}), it is clear that there
should be different solutions $\Phi$ associated with different values 
of the
free parameter $\Phi_0$. Intuitively, this parameter tunes
the size of the bosonic halo, since it defines the surface 
inside of which the scalar field plays a role.
\noindent In the recursion technique, a possible strategy could be to
impose $\Phi_0$ once and for all. Proceeding in that way, we found that the
solution did not converge properly. In fact, it is much more efficient
to choose arbitrarily
a point of coordinates $(r_0, \theta_0)$, and to impose step by step
that this point remains on the boundary; in other words, for each 
$n$, we
define $\Phi_0$ as $\Phi^{(n)}(r_0,\theta_0)$. For a given
$\theta_0$, the different possible choices of $r_0$ generate a
one--parameter family of solutions. 
However, only a finite range of $r_0$ values lead to a solution, 
from $r_{\rm min}=0$ for no bosonic halo, to
\beq 
r_{\rm max} \; = \; \pi \,
\sqrt{\displaystyle \frac{\lambda}{8 \, \pi \, G \, m^4}}
\; = \; \pi \, L
\eeq
for a pure scalar field configuration -- see eq.~(\ref{scale3})
or the alternative derivation in Appendix A. The choice of $\theta_0$
itself is irrelevant and we checked that any other choice gives the
same family of solutions.

\vskip 0.1cm
\noindent 
In summary, for a given baryonic density, the modified non--linear
Poisson can be solved by:
\begin{enumerate}
\item choosing an arbitrary $\theta_0$.
\item choosing a value $r_0$ reflecting the size of the scalar field halo.
\item defining a starting function $\Phi^{(0)}$.
\item integrating eq.~(\ref{eq:iterative}) recursively, 
with the help of equations
(\ref{decomposition}), (\ref{green_solution}) and (\ref{recomposition}).
\end{enumerate}
We show in Appendix B how to define a function $\Phi^{(0)}$ which is
close enough to the real solution in order to ensure fast convergence.

\begin{figure*}[!ht]
\centerline{
\epsfig{file=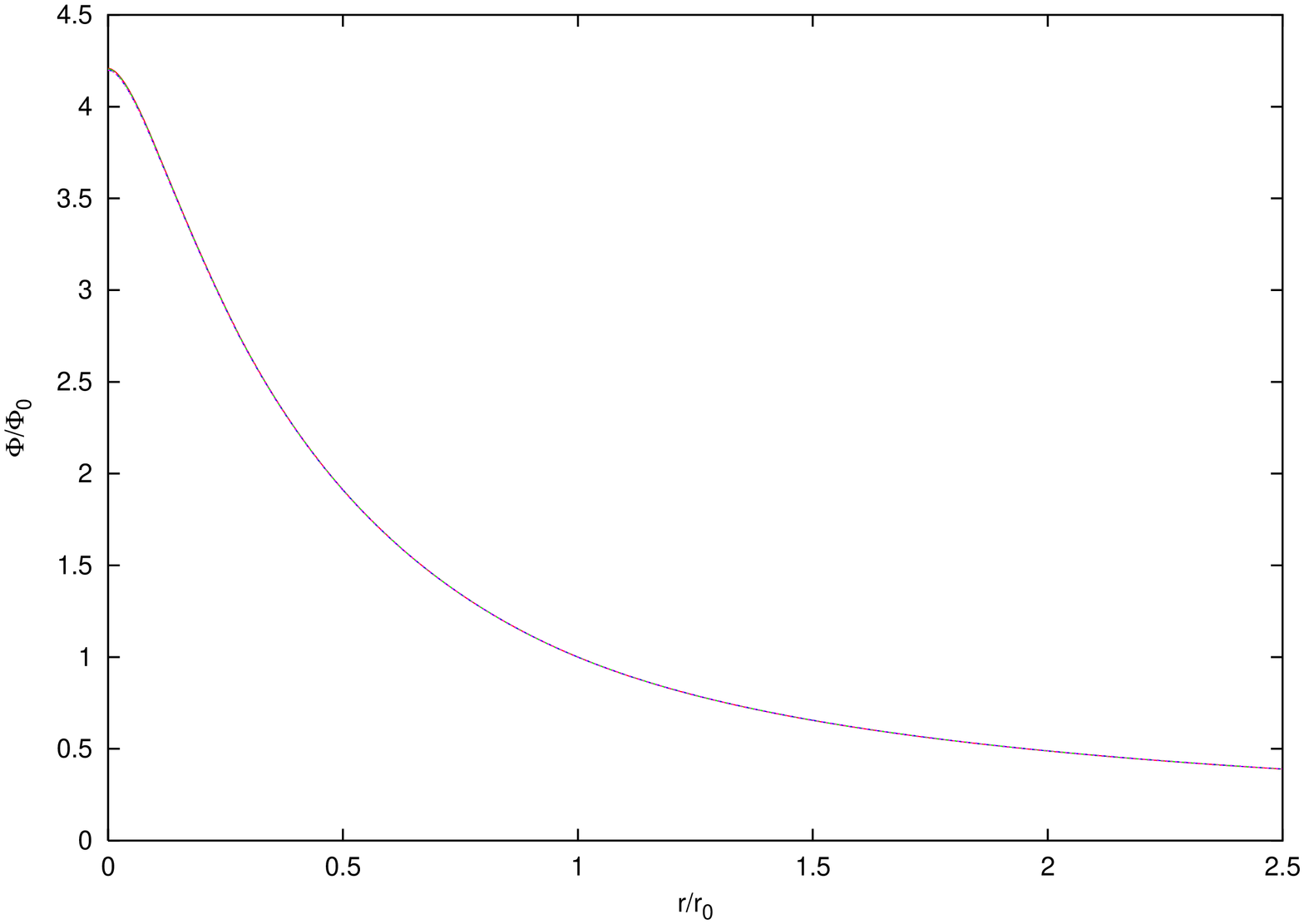,angle=270,width=0.75\textwidth}
}
\hspace{0.20\textwidth} 
\vspace{-0.55\textwidth}
$$
~~~~~~~~~~~~~~~~~~~~~~~~~~~~~~~~~~~~~\epsfig{file=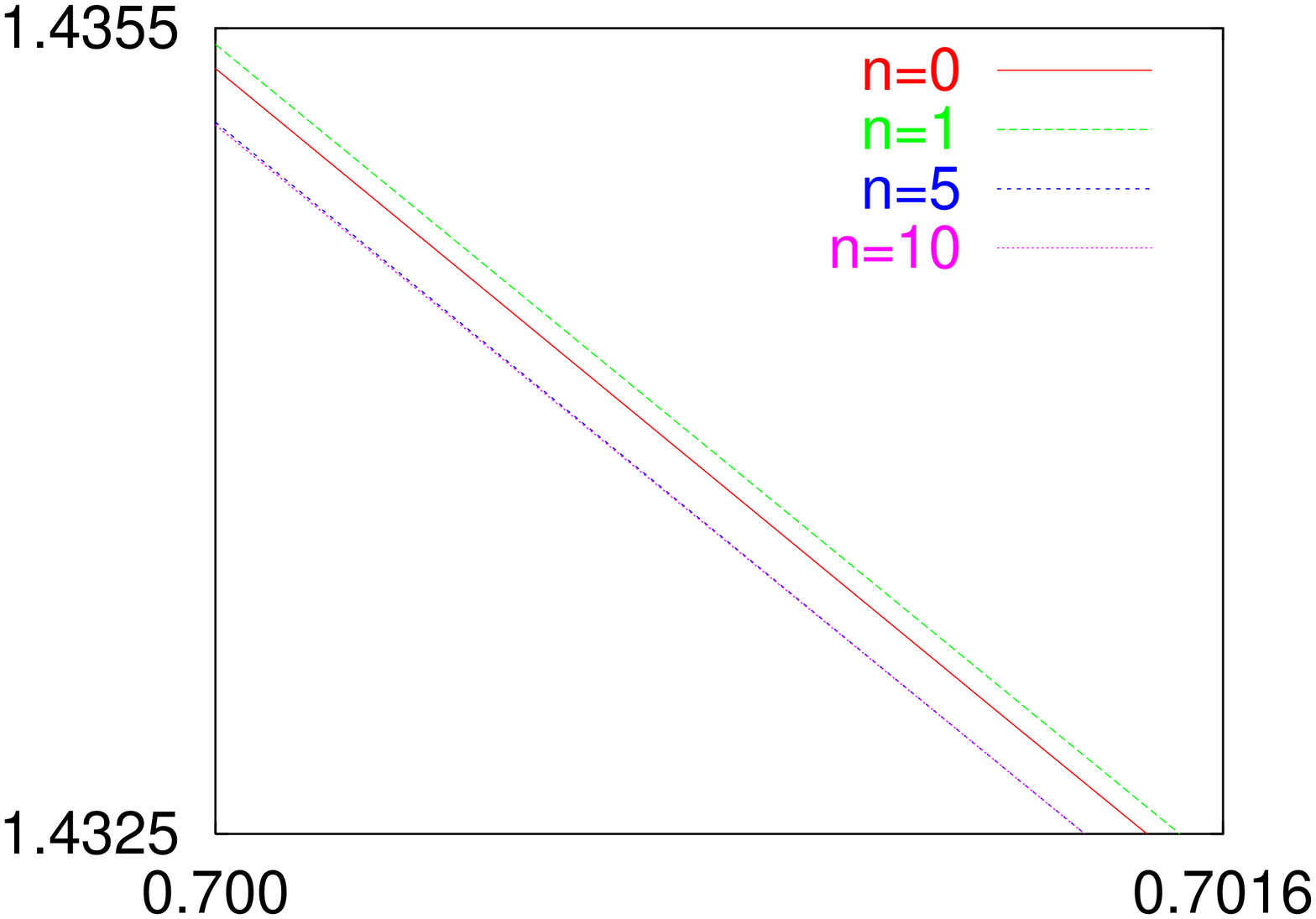,angle=270,width=0.35\textwidth}
$$
\vspace{0.25\textwidth}
\caption{
Gravitational potential $\Phi^{(n)}$ seen in the direction 
$\theta=\pi/2$ after $n=0$ (in red), 1 (in green), 5 (in blue) and 
10 
(in purple) recursions, with  $r_{\rm opt} = r_{\rm max}/ \pi$,
$r_0 = 2 r_{\rm max} / \pi$, $\theta_0 = \pi/2$.
The functions are seen to converge quickly.
}
\label{fig3.1}
\end{figure*}

\begin{figure*}[!ht]
\centerline{
\epsfig{file=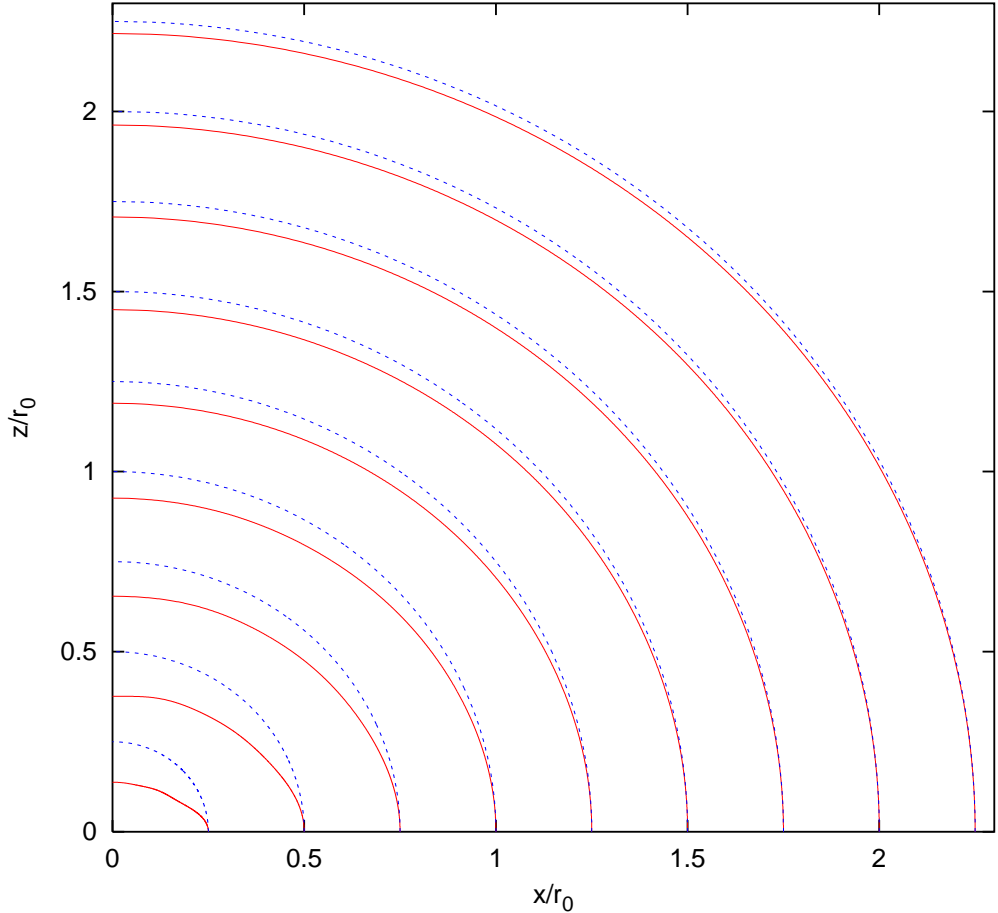,angle=270,width=0.75\textwidth}}
\vskip 0.1cm
\caption{
Equipotential lines generated by a disk-shaped density and a scalar field
halo (in red), 
compared to spherical equipotentials (in blue). This 
plot has been computed in the case ($r_{\rm opt} = r_{\max} / \pi$ , 
$r_0 = 2 r_{\rm max} / \pi$, $\theta_0=\pi/2$). 
The cartesian coordinate $(x, z)$ are such that the axis
of revolution of the galaxy corresponds to $x=0$.
}
\label{fig3.2}
\end{figure*}

\vskip 0.1cm
\noindent 
Let us illustrate this technique with a particular example,
based on a stellar disk plus a bosonic halo.
In the following, we will always treat the stellar disk
as a thin distribution with exponentially 
decreasing density. The optical radius is defined in such a way that it
encompasses 83\% of the total stellar mass, and the thickness of the 
disk is chosen to be twenty times smaller than its radius, so that
\beq
\rho_b(r,\theta) \propto \exp\left\{- 3.2 |\cos\theta| 
\frac{r}{r_{\rm opt}}\right\} \exp\left\{- 3.2 \gamma |\sin 
\theta|\frac{r}{r_{\rm opt}}\right\} \;\; ,
\label{baryon_distribution}
\eeq
where $\gamma=20$.
Let us choose a case where $r_0$ and $r_{\rm opt}$ are comparable,
so that baryons and bosons both have an influence, for instance: 
\beq
r_{\rm opt} 
= \frac{r_{\rm max}}{\pi}
\qquad
{\rm and} 
\qquad
(r_0, \theta_0) = ( 2 \frac{r_{\rm max}}{\pi}, \,  \frac{\pi}{2}) \, .
\eeq 
We plot on figure~\ref{fig3.1} the function 
\beq
\frac{\Phi^{(n)}(r, \pi/2)}{\Phi^{(n)}(r_0, \pi/2)}
\eeq
for $n=0, 1, 5, 10$.
In this example, the value $\Phi=\Phi_0$ is reached at 
$r = 2 \, r_{\rm opt}$ in the disk plane, and
$r= 1.85 \, r_{\rm opt}$ in the orthogonal direction.
The oblate form of the equipotentials is seen on figure~\ref{fig3.2}. \\
A good test of the recursive method is to pick up different values
of $\theta_0$, and to see whether there is always a value
$r_0(\theta_0)$ such that: $\Phi_0$
is always the same, and the solutions $\Phi(r, \theta)$ are exactly
identical. We checked this successfully on various examples.

\begin{figure*}[!ht]
\centerline{
\epsfig{file=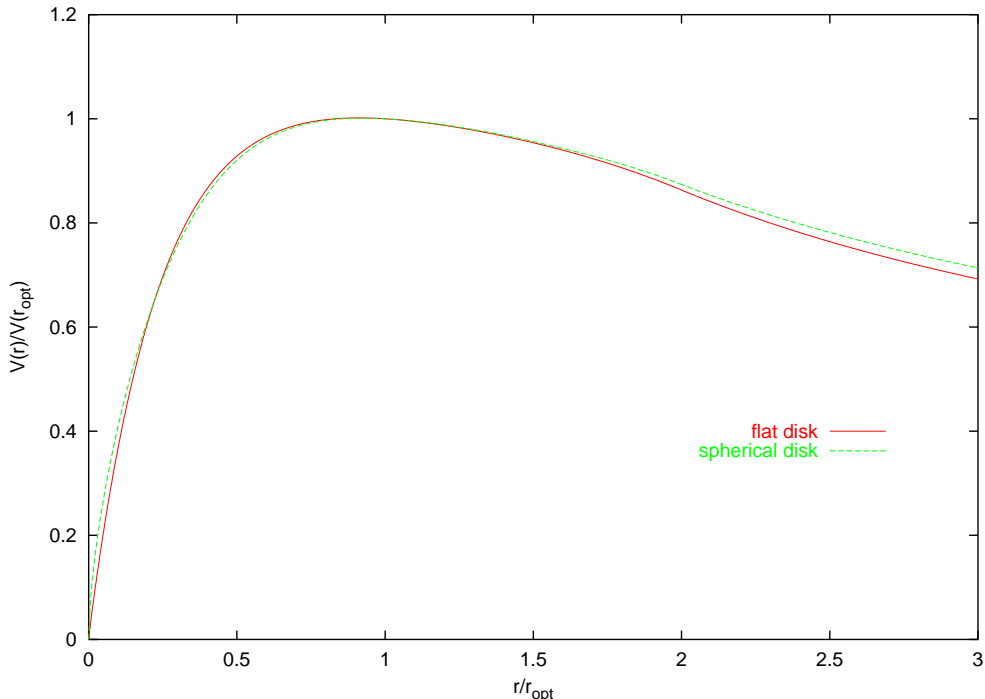,angle=270,width=0.75\textwidth}}
\vskip 0.1cm
\caption{
Rotation curve due to a disk-shaped density (in red) 
compared to that obtained in a spherical approximation (in green)
as in Paper~I. 
These two examples have been computed in the case ($r_{\rm opt} = 
r_{\rm max} / \pi$ , $r_0 = 2 r_{\rm opt}$, $\theta_0=\pi/2$).
The difference is seen to be fairly small.
}
\label{fig3.3}
\end{figure*}

\vskip 0.1cm
\noindent
The rotation curve can be deduced from the gravitational potential:
\beq
v^2=r \frac{\partial}{\partial r}\Phi(r, \theta= \pi/2)\,.
\eeq
\noindent On figure~\ref{fig3.3}, we compare the rotation curve 
obtained
following our method with the one calculated in the approximation
of Paper~I:
namely, replacing the thin disk by a spherical one,
with a density such that in absence of any halo, the rotation curve
along the stellar plane would be the same as with the true 
non-spherical disk. In presence of a halo, one can see that the 
difference becomes important only at large radius.

\section{The rotation curves of dwarf spirals}
\label{sec:rotation_curves}

\noindent
Dwarf spiral galaxies are known to be completely dominated
by dark matter at all radii. Usual CDM models fail to reproduce
the rotation curves of those systems. The purpose of this analysis 
is to investigate whether a self--interacting massive scalar field 
halo is able to reproduce such rotation curves. Therefore, we will
first scrutinize the typical dwarf spiral galaxy
DDO~154 that has been thoroughly studied -- see for example
the observations by \cite{hoffman} and \cite{carignan}. Because
it is isolated and therefore seems to be protected against
any external influence, this dwarf spiral features a prototypical
example for our study. Its HI gas contribution is well measured
and follows the distribution
\beq
\rho_{\rm gas} \left( r , \theta \right) \; = \;
\rho_{\rm gas}^{c} \, \exp \left\{
- 0.8 \, |\cos \theta| \, {\displaystyle \frac{r}{r_{\rm opt}}} \right\}
\, \exp \left\{ - 5 \, |\sin \theta| \,
{\displaystyle \frac{r}{r_{\rm opt}}} \right\} \;\; .
\label{eq:gas_density}
\eeq
Its optical radius $r_{\rm opt}$ is equal to 1.4 kpc.
The contribution of its stars is visible and therefore well--known,
with a density distribution given by relation~(\ref{baryon_distribution}).
Both stars and gas account for a small fraction of the observed
circular velocity.

\begin{figure*}[!ht]
\centerline{\epsfig{file=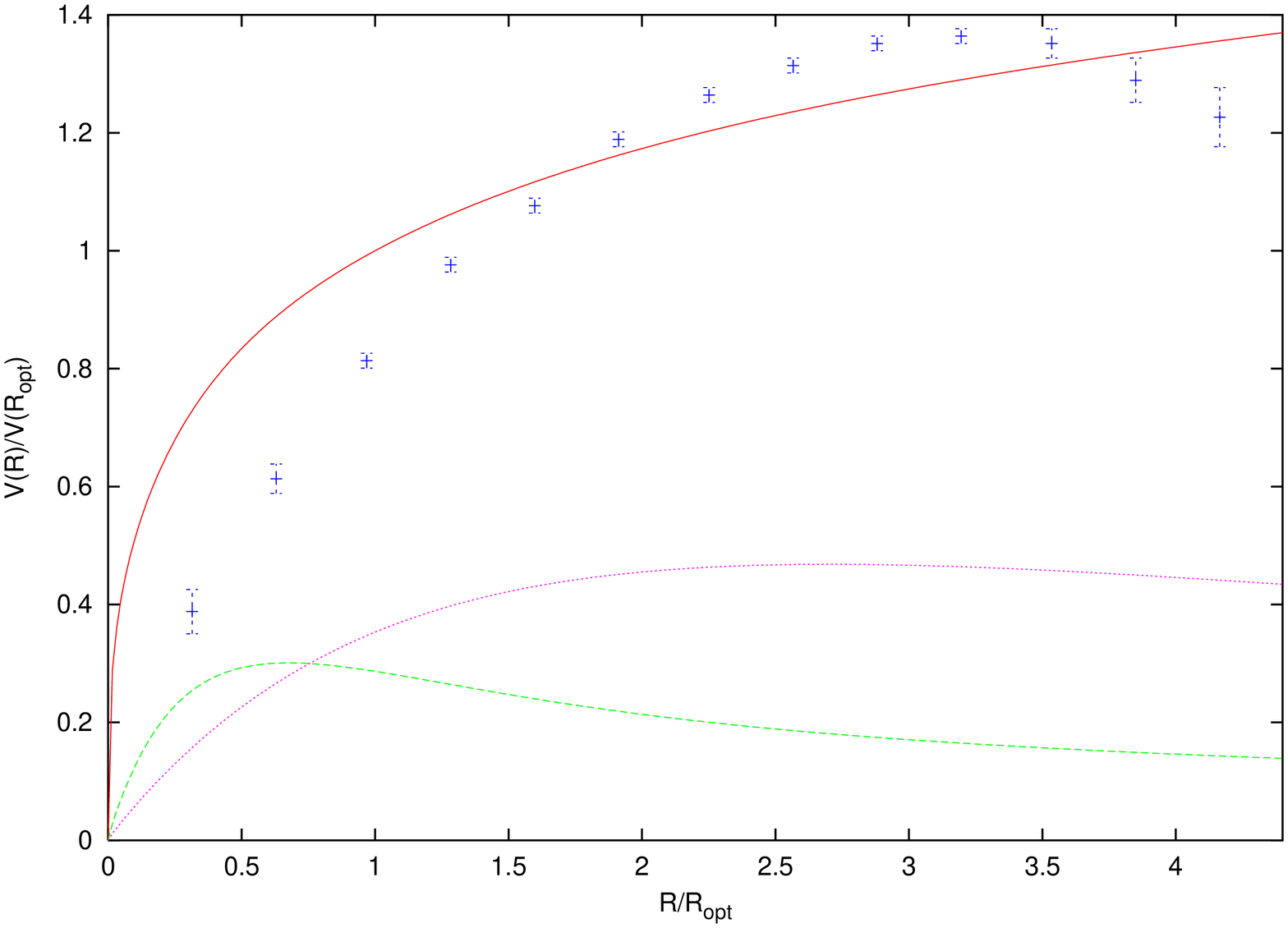,height=70mm,angle=270}
\epsfig{file=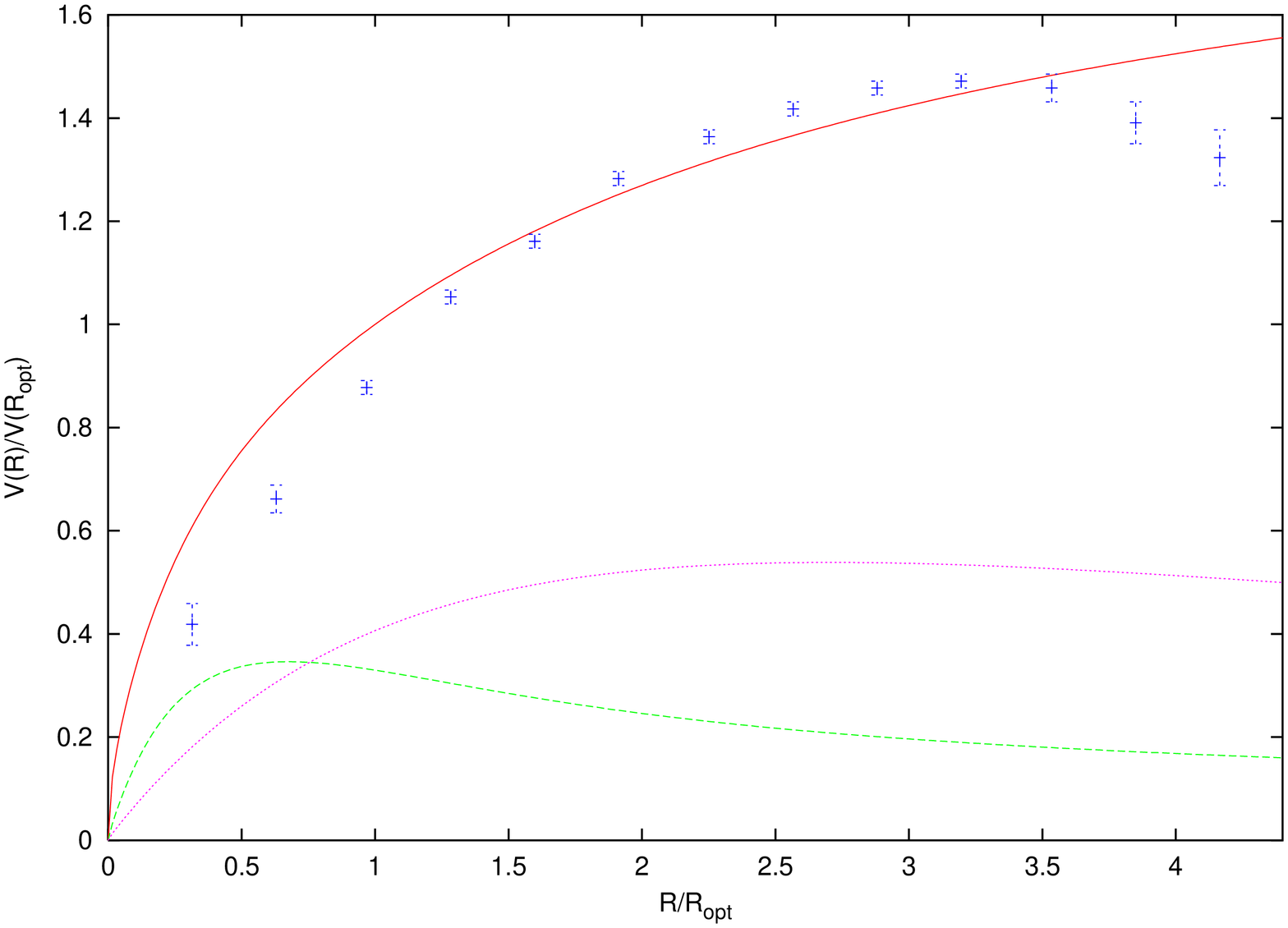,height=70mm,angle=270}}\par
\centerline{\epsfig{file=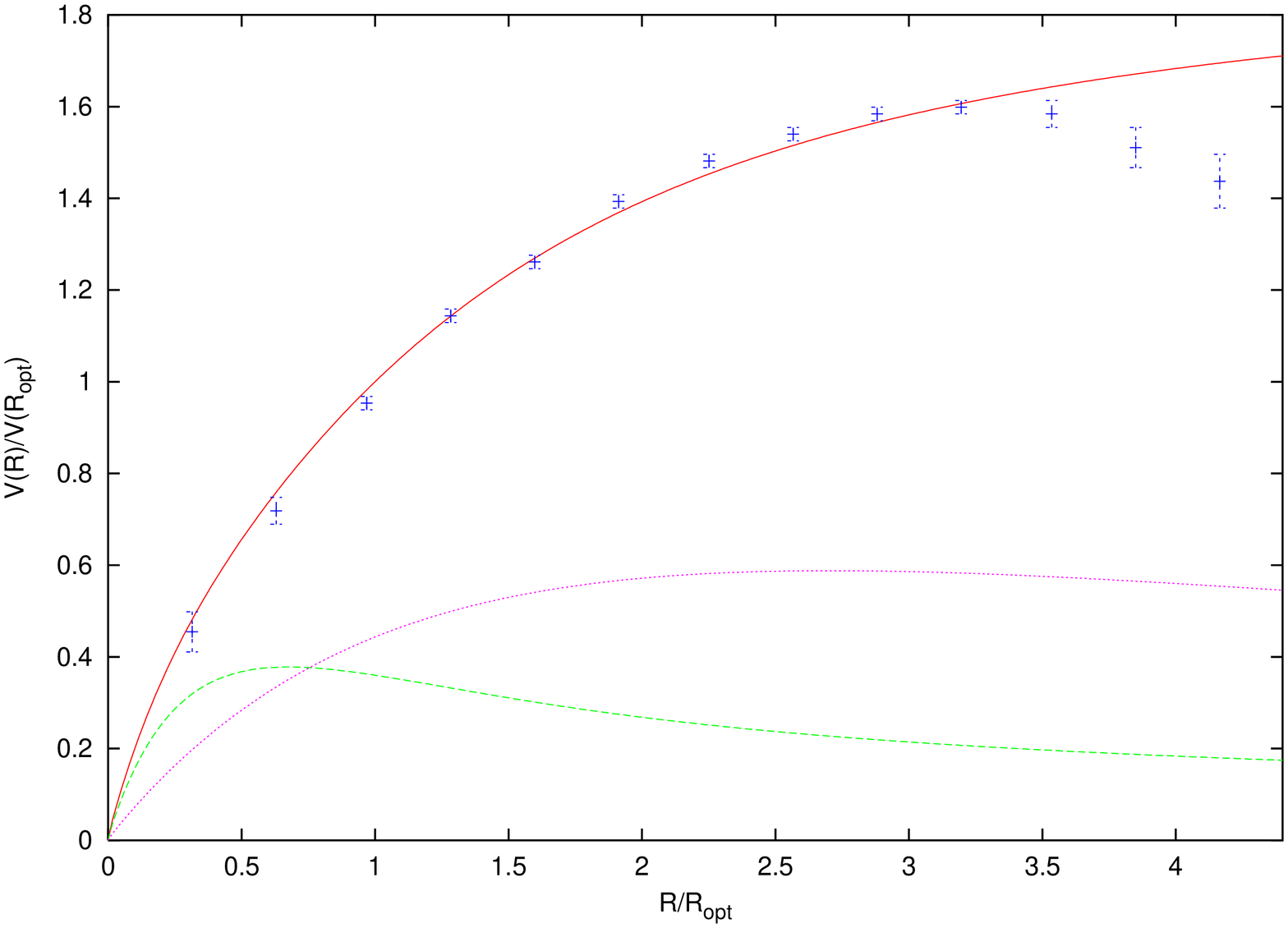,height=70mm,angle=270}
\epsfig{file=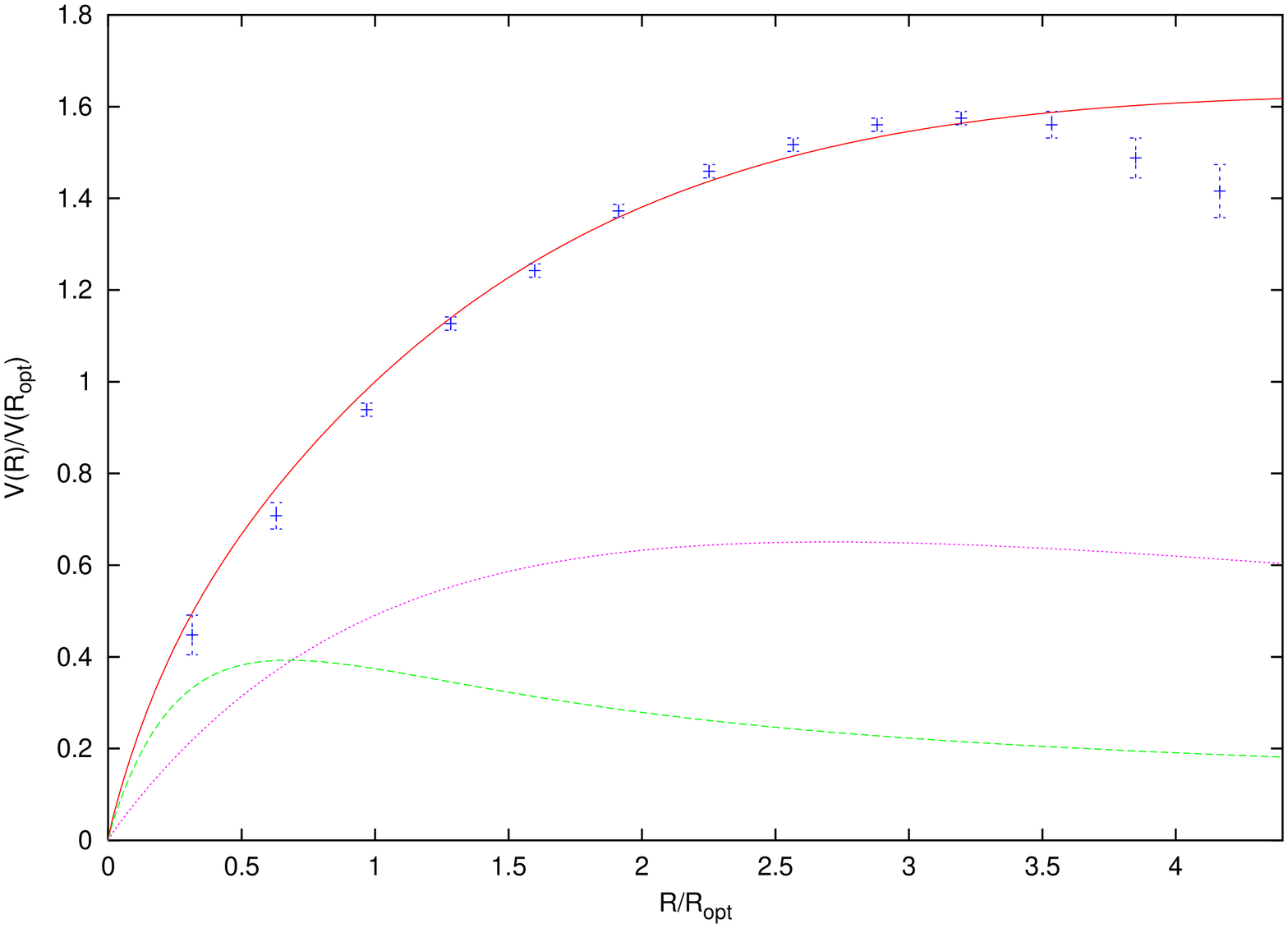,height=70mm,angle=270}}
\vskip 1.cm
\caption{From left to right and top to bottom, the best fit
of the DDO~154 rotation curve (in red) is featured as well as
the observed stellar (in green) and gas (in purple) densities.
A dark matter component has been added and various profiles
have been assumed~: Moore's density (a), NFW profile (b),
isothermal halo (c) and Burkert's phenomenological
distribution (d).}
\label{fig:fig4}
\end{figure*}

\vskip 0.1cm
\noindent
In order to compare the various dark matter models with the
observations of DDO~154, we have performed a $\chi^{2}$ test
on the 13 data points from ref.~\cite{carignan}.
To commence, we have considered models in which the approximate
real density of stars and gas has been assumed with
$\rho_{\rm gas}^{c} \approx 0.15 \, \rho_{\rm stars}^{c}$ and
$v_{\rm stars}(r_{\rm opt}) \approx 0.3$ -- $0.4 \;
v_{\rm tot}(r_{\rm opt})$ -- where $v_{\rm stars}$ is the
stellar contribution to the rotation velocity $v_{\rm tot}$.
The precise value of the ratio
$v_{\rm stars}(r_{\rm opt}) / v_{\rm tot}(r_{\rm opt})$
is unknown and has been adjusted here in order to provide the
best fit.
On top of stars and gas, a dark matter component is added with
a density profile that depends on the model at stake.
%
The Moore's model \cite{moore} is featured in the panel (a) of
fig.~\ref{fig:fig4} and corresponds to the spherical symmetric
density
\begin{equation}
\rho_{\rm M}(r) \; = \; \rho_{\rm M}^{c} \; \left\{ {\displaystyle
\frac{r_{\rm opt}^{3}}{r^{1.5} \, \left( r + r_s \right)^{1.5}}}
\right\}
\;\; ,
\end{equation}
where $r_s$ is a scale radius parameter that is also adjusted
in the fit.
Recent CDM N--body simulations point towards such a profile.
In the case of DDO~154, the best fit corresponds to
$\rho_{\rm M}^{c} \, \left( r_{\rm opt} / r_s \right)^{1.5}
\approx 0.07 \, \rho_{\rm stars}^{c}$ and to very large values
of the scale radius $r_s$. The $\chi^{2}$ value is found to be
approximatively equal to 600 for 10 degrees of freedom.
As previously mentioned, Moore's model where the density
diverges like $r^{-1.5}$ in the central region fails to
account for the dark matter distribution inside DDO~154.
%
Then, we have tested a NFW spherical density profile \cite{NFW}
where
\begin{equation}
\rho_{\rm NFW}(r) \; = \; \rho_{\rm NFW}^{c} \;
\left\{ {\displaystyle
\frac{r_{\rm opt}^{3}}{r \, \left( r + r_s \right)^{2}}}
\right\} \;\; .
\end{equation}
Such a distribution peaks at the center and has also been found
to naturally arise in N--body numerical simulations of neutralino
dark matter. We find that the best $\chi^{2}$ lies around 200
when $r_s \approx 9 \, r_{\rm opt}$ and
$\rho_{\rm NFW}^{c} \approx 6 \, \rho_{\rm stars}^{c}$ -- see
panel (b) of fig.~\ref{fig:fig4}.
%
We have also considered an isothermal spherical halo \cite{iso}
with
\begin{equation}
\rho_{\rm iso}(r) \; = \; \rho_{\rm iso}^{c} \;
\left\{ {\displaystyle
\frac{r_{\rm opt}^{2}}{r^2 + r_s^2}} \right\} \;\; .
\end{equation}
This density has been introduced in order to account for flat
rotation curves. In the case of DDO~154 where the circular speed
starts to decrease beyond 4.5 kpc, the best fit is obtained for
$r_s \approx 1.2 \, r_{\rm opt}$ and
$\rho_{\rm iso}^{c} \approx 0.15 \, \rho_{\rm stars}^{c}$.
The corresponding $\chi^{2}$ is now far more better with a value
$\sim$ 55 -- see panel (c) of fig.~\ref{fig:fig4}.
%
Finally, a Burkert spherical distribution \cite{Burkert}
\begin{equation}
\rho_{\rm B}(r) \; = \; \rho_{\rm B}^{c} \
\left\{ {\displaystyle
\frac{r_{\rm opt}^{3}}
{\left( r + r_s \right) \, \left( r^2 + r_s^2 \right)}}
\right\} \;\; ,
\end{equation}
has been considered. This density law has a core radius of size
$r_{s}$ -- just like the isothermal halo -- and converges at large
distances towards a Moore or a NFW profile. It has been introduced
as a phenomenological explanation of the rotation curves of
dwarf galaxies \cite{Salucci}. The best parameters are then
$r_s \approx 1.9 \, r_{\rm opt}$ and
$\rho_{\rm B}^{c} \approx 6 \, \rho_{\rm stars}^{c}$, leading to
a best $\chi^{2} \sim 45$ -- see panel (d) of fig.~\ref{fig:fig4}.
%
At this stage, we reach the conclusion that neutralino dark matter
-- should it collapse according to the N--body numerical simulations
\`a la Moore or NFW -- is too much peaked at the center of DDO~154
and does not account for the rotation curve in that region. The fact
that these species fail to reproduce the inner dynamics of a system
known to be saturated by dark matter is definitely a problem.
The isothermal and Burkert halos provide a better agreement with
the data but are not consistent with the decrease observed beyond
4.5 kpc.

\begin{figure*}[!ht]
\centerline{
\epsfig{file=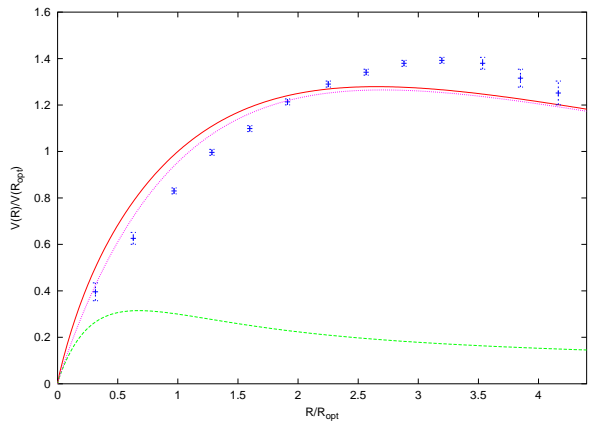,height=80mm,angle=270}}
\vskip 1.cm
\caption{The DDO~154 rotation curve (in red) is fitted
with the observed stellar density (in green) while the
gas distribution (in purple) has been artificially
enhanced with respect to the observed HI by a rescaling
factor.}
\label{fig:fig5}
\end{figure*}

\vskip0.1cm
\noindent
We have then investigated a slightly different idea. Following
Pfenniger and Combes \cite{COMBES1},
the dark matter inside galaxies would consist of pure molecular
hydrogen $H_{2}$ -- so cold that it would have gone undetected so
far. The formation of stars in the inner parts and its concomitant UV
light production would have turned part of the $H_{2}$ into detectable
HI. The distribution of this hidden $H_{2}$ component could
be derived in the case of DDO~154 from its observed rotation curve.
We will nevertheless adopt the opposite point of view since
our aim is to derive -- and not to start from -- the circular speed.
We have therefore artificially
rescaled the observed gas density~(\ref{eq:gas_density}) by a
homogeneous overall factor. The best fit featured in
fig.~\ref{fig:fig5} corresponds to
$\rho_{\rm gas}^{c} \; = \; \rho_{\rm stars}^{c}$ and leads to
a best $\chi^{2}$ of $\sim$ 500 which is not particularly exciting.
In the case of the models of fig.~\ref{fig:fig4}, the addition of
such a cold gas component does not improve the goodness of our fits.

\begin{figure*}[!ht]
\centerline{
\epsfig{file=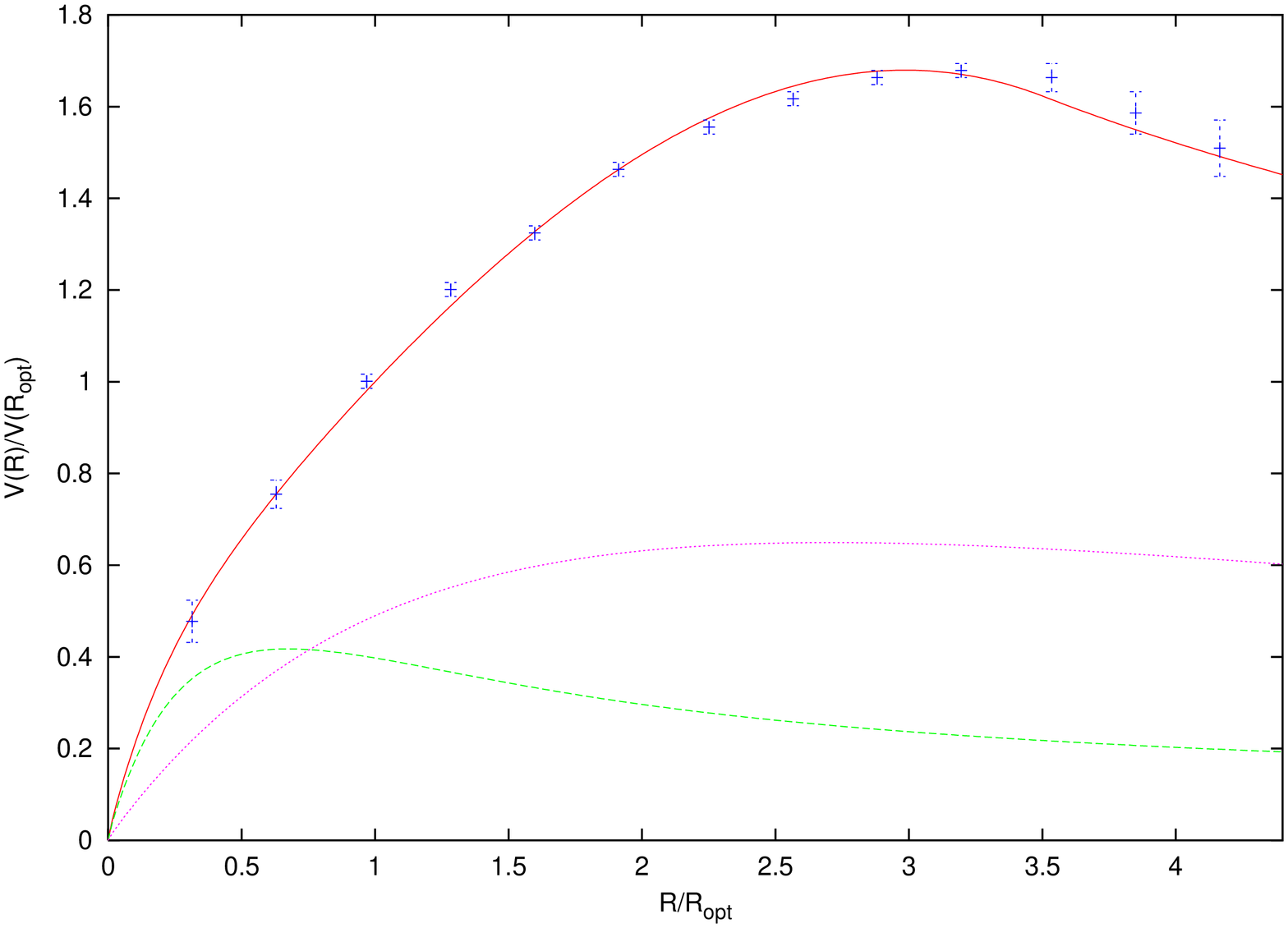,height=80mm,angle=270}
\epsfig{file=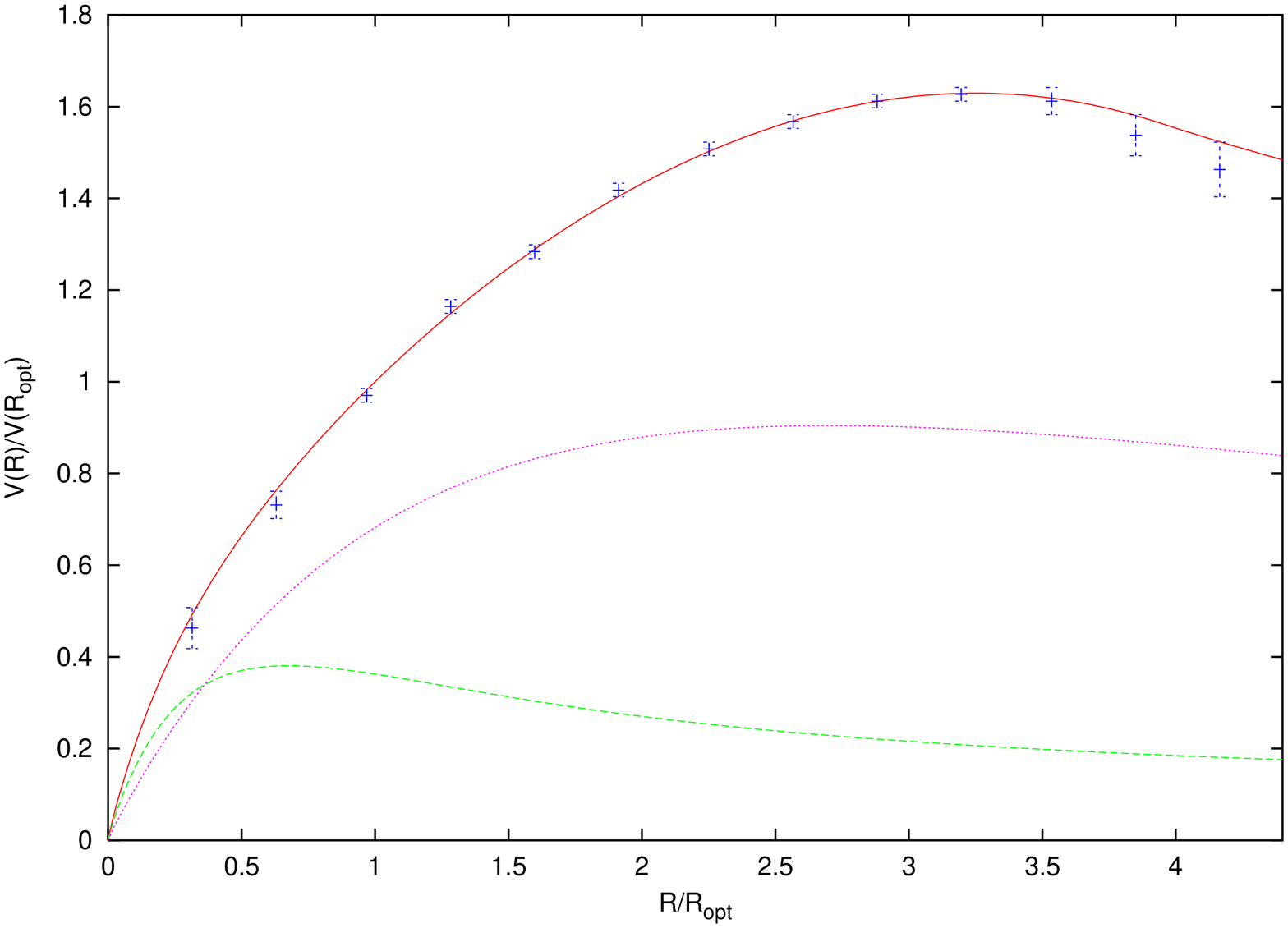,height=80mm,angle=270}}
\vskip 1.cm
\caption{The best fit of the DDO~154 rotation curve (in red)
is presented with the observed stellar density (in green)
and gas distribution (in purple).
In the left panel, a self--interacting bosonic halo is
assumed with $m^{4} / \lambda \approx 75$ eV$^4$ together
with the observed gas density profile.
In the right panel, the gas component has been rescaled
in order to improve the goodness of fit and a value of
$m^{4} / \lambda \approx 50$ eV$^4$ is derived.}
\label{fig:fig6}
\end{figure*}

\vskip0.1cm
\noindent
Finally, we have assumed the presence of a self--interacting
bosonic halo and applied the recursion method
discussed in section~\ref{sec:method}. The left plot
of fig.~\ref{fig:fig6} corresponds to stellar and gas
populations as observed while a value of
$m^{4} / \lambda \approx 75$ eV$^4$ provides a best $\chi^{2}$
of 16. The agreement with the measured rotation curve is quite
good. Notice that the bosonic halo dominates completely the
inner dynamics beyond $\sim$ 0.5 kpc.
More impressive is the right plot of fig.~\ref{fig:fig6} where
the gas distribution has now been rescaled in order to improve the
goodness of fit. A best $\chi^{2}$ of $\sim 7$ is reached for
$\rho_{\rm gas}^{c} \approx 0.35 \, \rho_{\rm stars}^{c}$
and a value of $m^{4} / \lambda \approx 50$ eV$^4$.

\begin{figure*}[!ht]
\centerline{
\epsfig{file=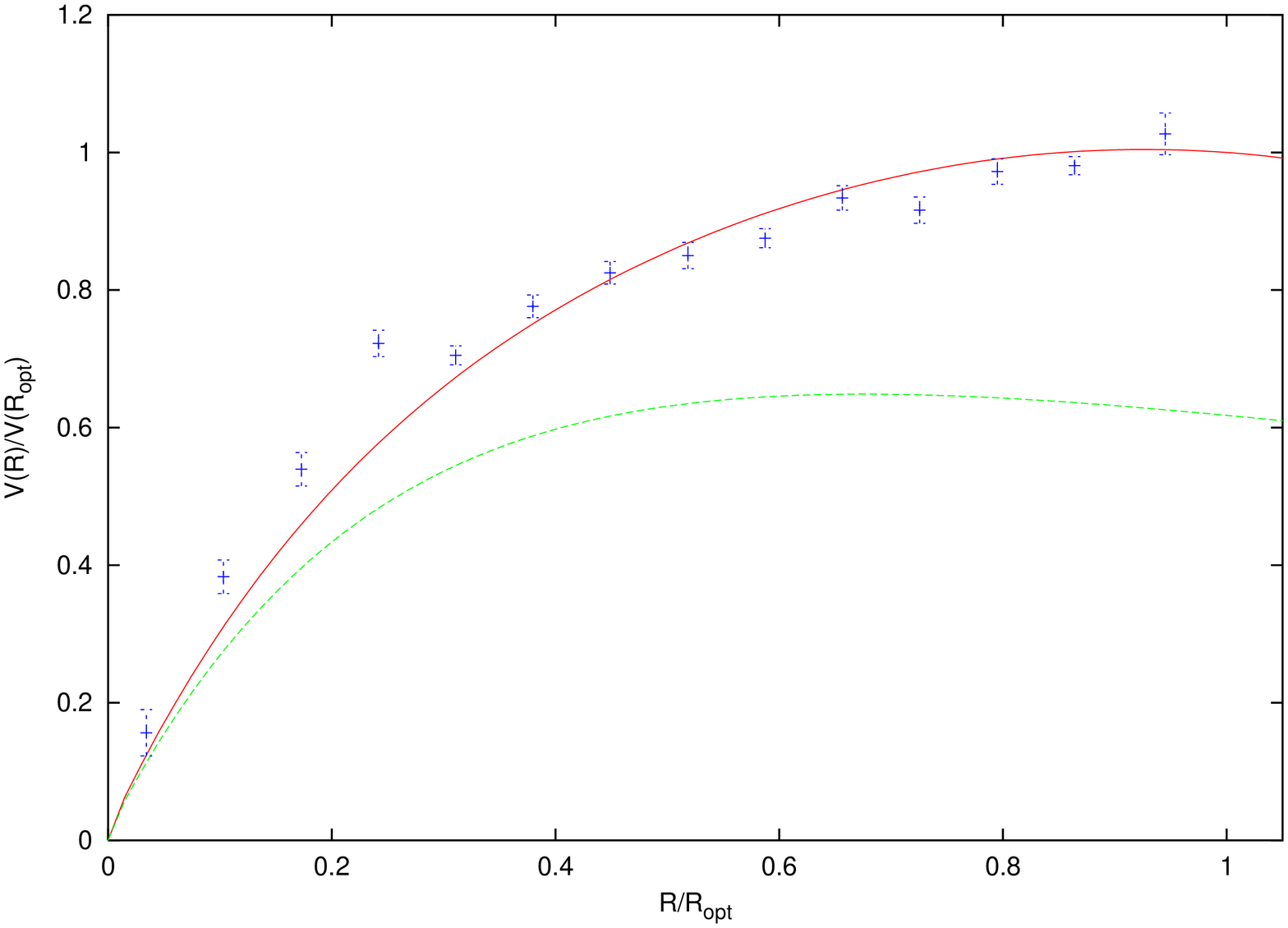,height=70mm,angle=270}
\epsfig{file=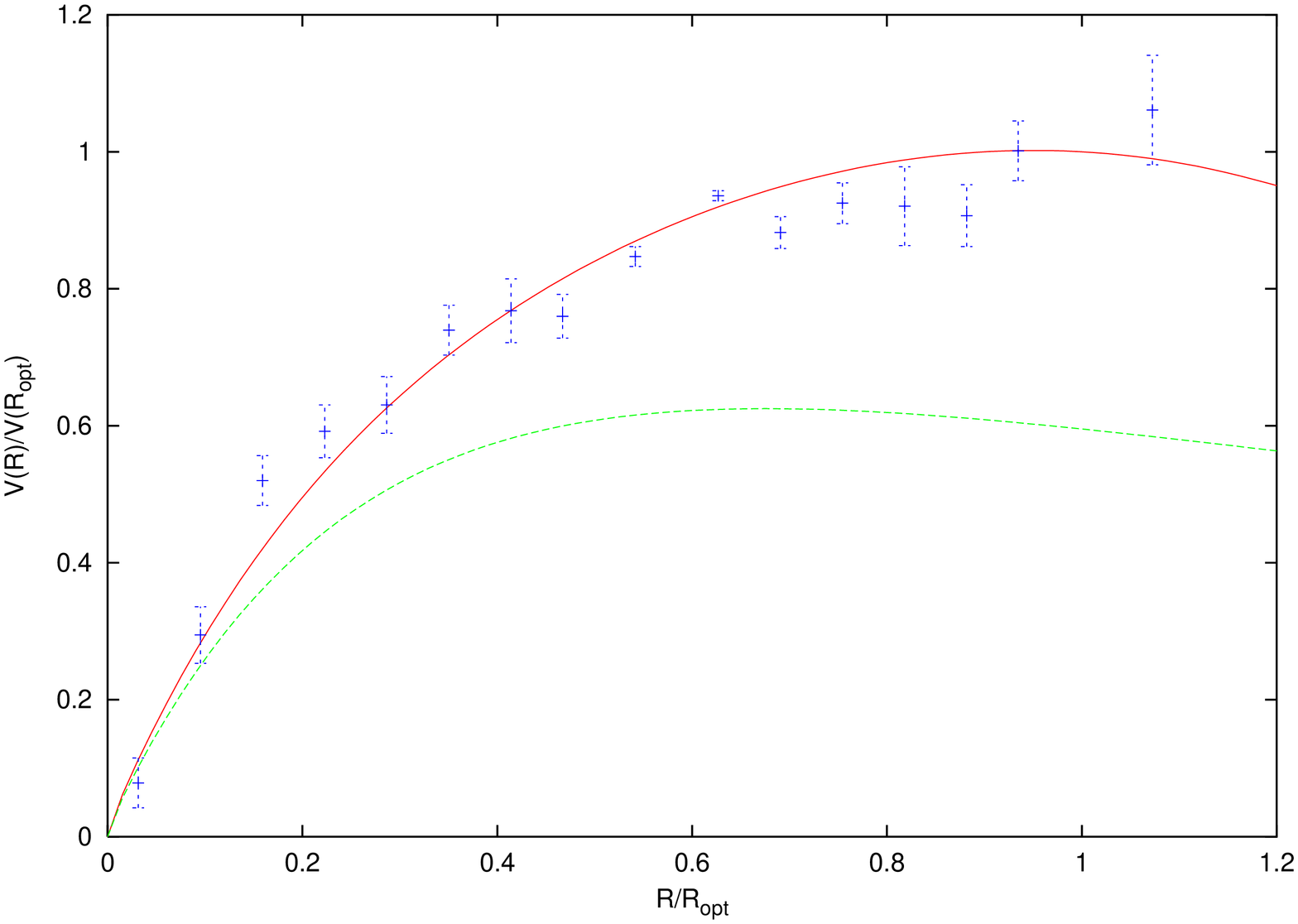,height=70mm,angle=270}}
\centerline{
\epsfig{file=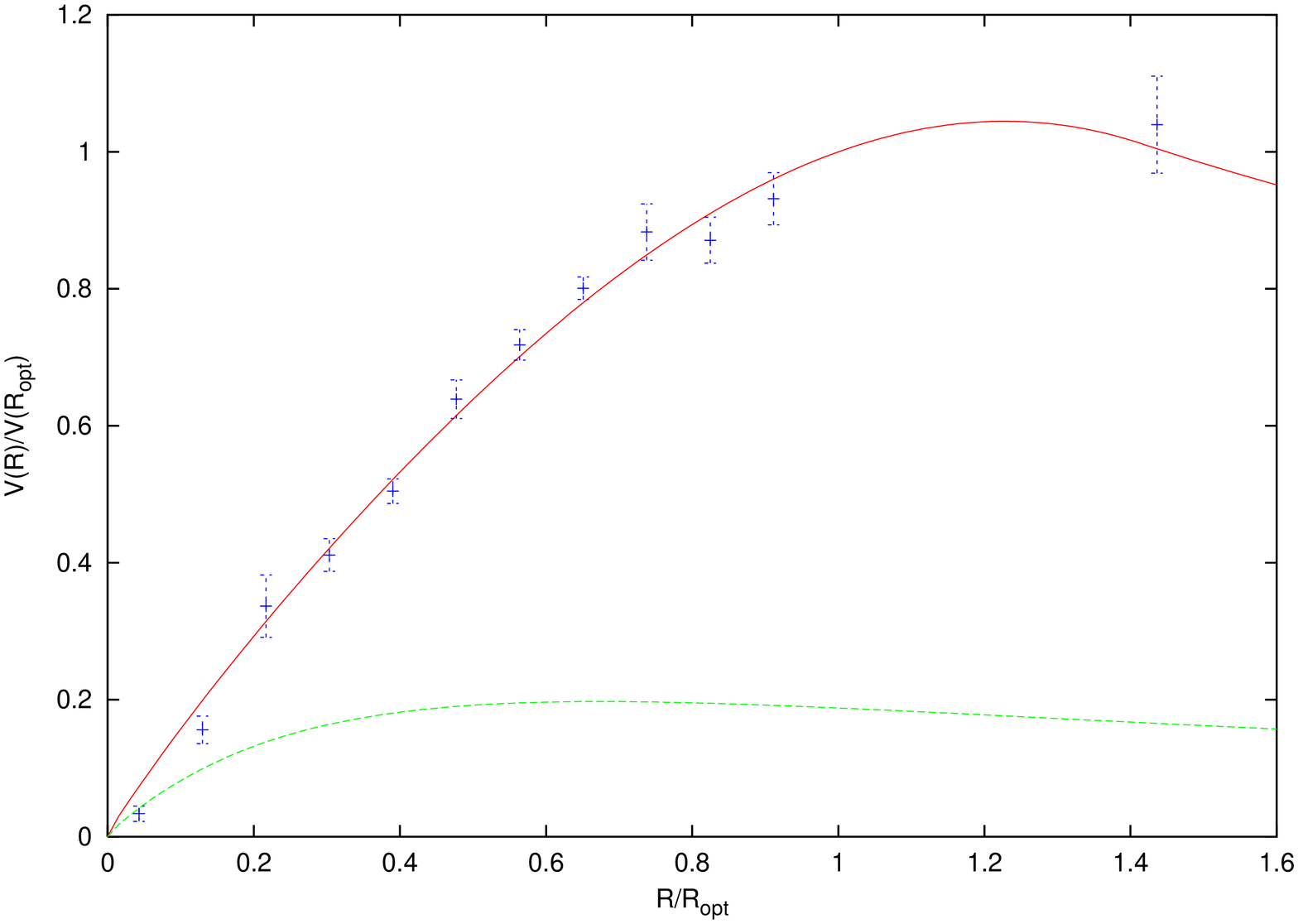,height=70mm,angle=270}
\epsfig{file=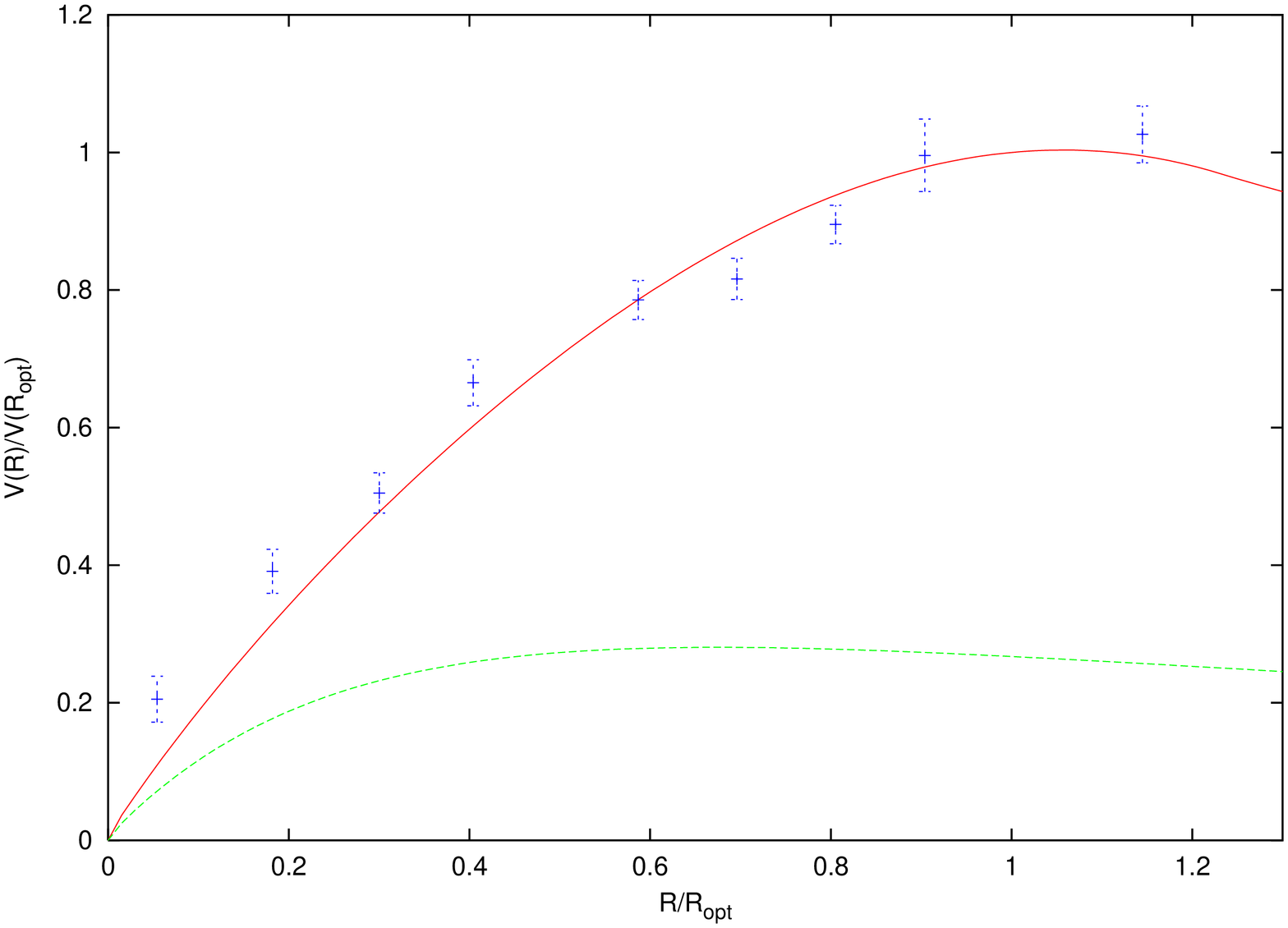,height=70mm,angle=270}}
\centerline{
\epsfig{file=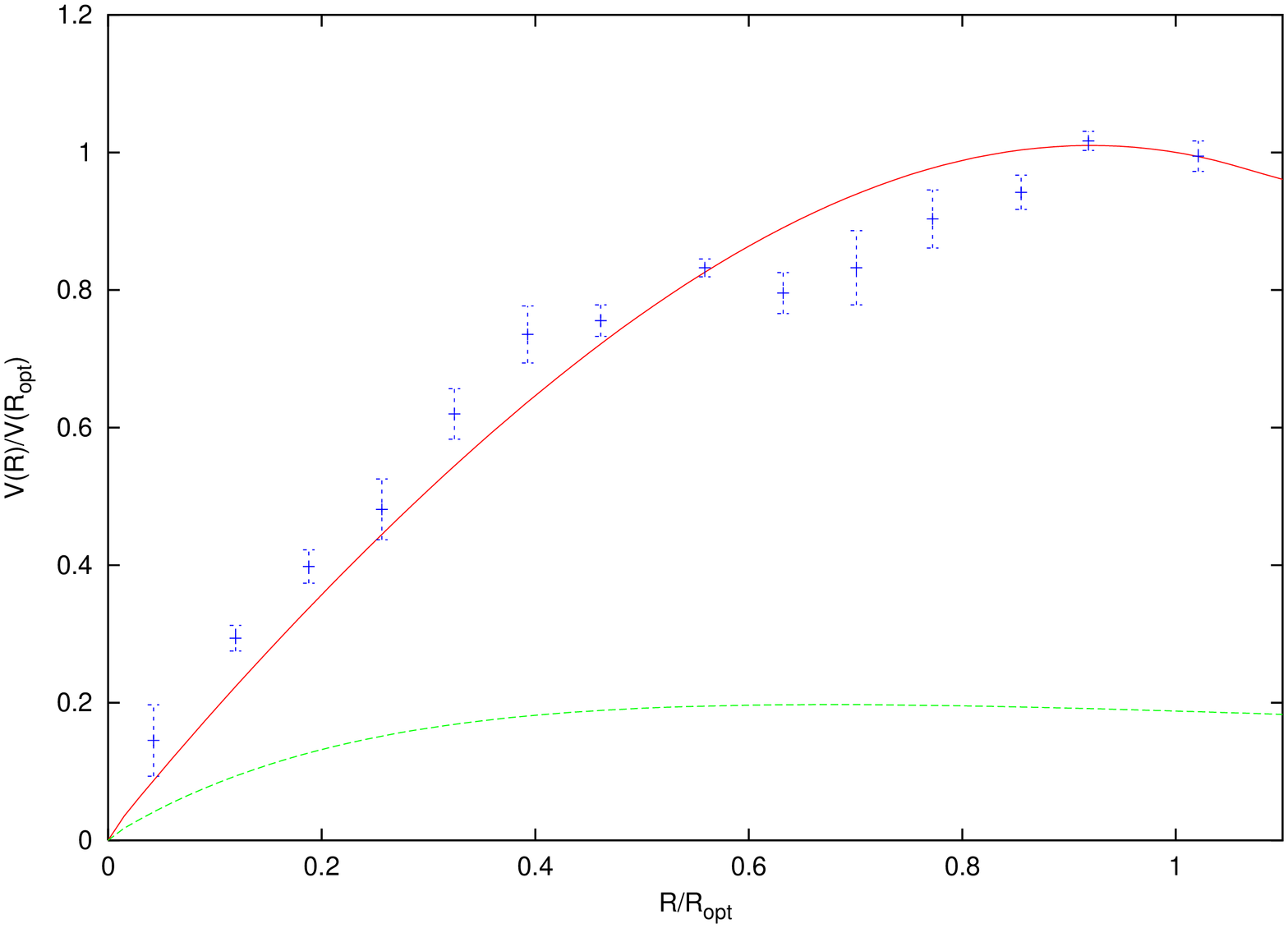,height=70mm,angle=270}
\epsfig{file=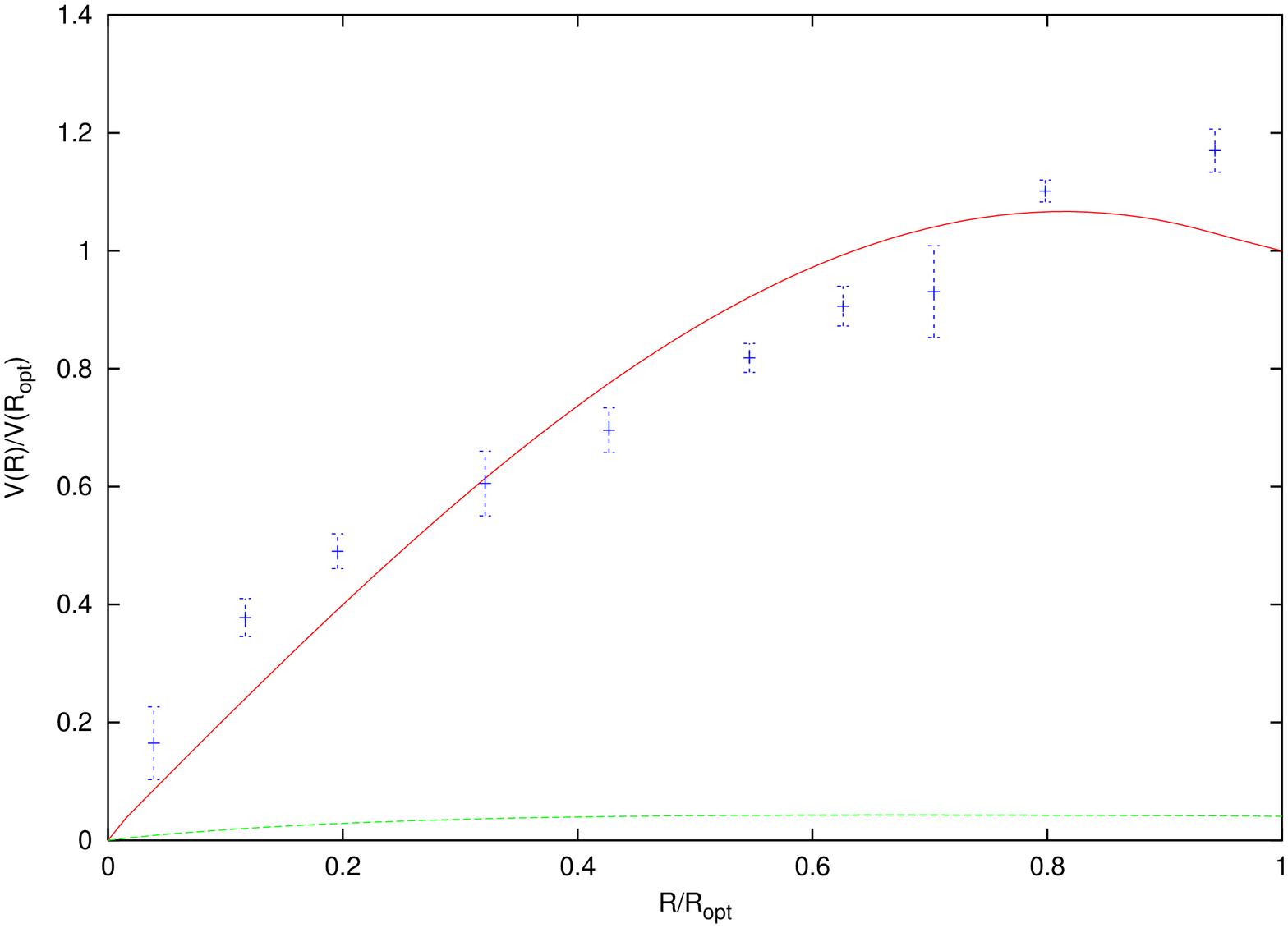,height=70mm,angle=270}}
\vskip 1.cm
\caption{From left to right and top to bottom, the best fit
(in red) of the rotation curves of various spiral galaxies
with increasing optical radii. The green lines stand for
the stellar contributions. The various panels respectively
correspond to
(a) N7339      ($r_{\rm opt}$ = 4.8 kpc) --
(b) M--3--1042 ($r_{\rm opt}$ = 4.8 kpc) --
(c) N755       ($r_{\rm opt}$ = 4.8 kpc) --
(d) 116--G12   ($r_{\rm opt}$ = 5.4 kpc) --
(e) 563--G14   ($r_{\rm opt}$ = 6.4 kpc) --
(f) 545--G5    ($r_{\rm opt}$ = 7.7 kpc).}
\label{fig:fig7}
\end{figure*}

\vskip0.1cm
\noindent
Beside the prototypical example of DDO~154, we have
analyzed a set \cite{Salucci} of small and medium size
spiral systems for which measurements of the rotation
curve are of high quality.
These galaxies have been selected on the requirement that
they have no bulge, very little HI -- if any -- and a dominant
stellar disk that accounts for the dynamics in the central
region.
They are also dominated by dark matter as is clear from
fig.~\ref{fig:fig7}.
A self--interacting bosonic halo has been assumed with
$m^{4} / \lambda \approx 50$ eV$^4$. Because of the presence
of wiggles in the rotation curves -- presumably related
to spiral arms inside the disks -- the best $\chi^{2}$ value
becomes meaningless. The qualitative agreement is nevertheless
correct except in the case of 545--G5 where the optical radius
is $r_{\rm opt}$ = 7.7 kpc. Because the mass $m$ and the
coupling $\lambda$ define a unique scale of $\sim$ 2.3 kpc
-- see relation~(\ref{scale2}) -- the Bose condensate does
not extend enough to account for the dark matter inside large
systems. A single self--interacting bosonic halo fails to
reproduce at the same time the dark matter inside light
and massive spirals.
A possible solution lies in the existence of several small
bosonic condensates or clumps inside the halos of large
galaxies whereas a single condensate would account for the
dark matter of dwarf spirals such as DDO~154. We have
concentrated on dwarf spiral galaxies for which neutralinos
seem to be actually in trouble. The case of several bosonic
clumps is beyond the scope of this work and will be
investigated elsewhere.

\section{The solar system}
\label{sec:solar_system}

\noindent
As long as we were interested in the inner dynamics of galactic
systems, the baryonic density $\rb$ in equations~(\ref{poisson1})
and (\ref{poisson2}) was implicitly averaged over distances of order
a few pc and behaved smoothly. If the stellar population is now
made of point--like particles with mass $M_{i}$, the gravitational
potential $\Phi$ varies according to
\beq
\Delta \Phi \, + \,
8 \, \pi \, G \; {\displaystyle  \frac{m^{4}}{\lambda}} \,
\left( \Phi - \Phi_{0} \right) \, {\cal H} \left( \Phi_{0} - \Phi \right)
\; = \; 4 \, \pi \, G \;
{\displaystyle \sum_{i}} \, M_{i} \,
\delta^{3} \left( \vec{r} - \vec{r_{i}} \right) \;\; .
\label{poisson3}
\eeq
Outside the Bose condensate, the usual Poisson equation is recovered
so that the gravitational attraction of a star -- say the Sun -- is
not modified with respect to the conventional situation.
Slightly different is the case where the Sun lies inside the region
where the field $\phi$ extends. Intuitively, the scalar field is expected
to be attracted by the solar gravity and to concentrate around the Sun whose
gravity should consequently be strengthened. In order to investigate that
effect, we first notice that the potential difference $\Phi - \Phi_{0}$
is a linear function of the sources within the Bose condensate. The
contribution $\Phi_{\odot}$ of the sun to the potential difference
$\Phi - \Phi_{0}$ satisfies the modified Poisson equation
\beq
\Delta \Phi_{\odot} \, + \,
8 \, \pi \, G \; {\displaystyle  \frac{m^{4}}{\lambda}} \, \Phi_{\odot}
\; = \; 4 \, \pi \, G \; \rho_{\odot} \;\; ,
\label{poisson_sun}
\eeq
with the condition that it must vanish on the boundaries of the Bose
condensate. If the solar system is well embedded inside the latter --
below a depth well in excess of a few AU -- the surface of the condensate
is so far that we may just require that $\Phi_{\odot}$ vanishes at infinity.

\vskip 0.1cm 
\noindent
Assuming in addition that the solar density $\rho_{\odot}$ has an
isotropic distribution, the solution of eq.~(\ref{poisson_sun})
readily obtains in terms of the spherical Bessel functions
$j_{0}(z) = {\sin}z / z$ and $n_{0}(z) = - \, {\cos}z / z$ as
explained in Appendix B
\beq
\Phi_{\odot}(r) \; = \; - \, 4 \, \pi \, G \;
{\displaystyle \frac{L^{3}}{r}} \; \left\{
{\cos}z \, {\displaystyle \int_{0}^{\displaystyle z}}
\rho_{\odot}(u) \, u \, {\sin}u \, du \; + \;
{\sin}z \, {\displaystyle \int_{\displaystyle z}^{+ \infty}}
\rho_{\odot}(u) \, u \, {\cos}u \, du \right\} \;\; .
\eeq
The dimensionless radial coordinate $z$ is defined as the ratio
$r / L$ where the typical scale $L$ has already been defined in
section~\ref{sec:basic_equations}. Relations~(\ref{scale1}) and
(\ref{scale2}) imply that $L$ exceeds the solar radius $R_{\odot}$
by some ten to eleven orders of magnitude. The gravitational potential
which the Sun generates with the help of the scalar field $\phi$
simplifies into
\beq
\Phi_{\odot}(r) \; = \; - \,
{\displaystyle \frac{G \, M_{\odot}}{r}} \,
\cos \left( {\displaystyle {r}/{L}} \right) \;\; .
\eeq
Because of our assumption as regards the boundary condition -- which
we placed at infinity -- this relation may be safely used only for
distances $r \ll L$. Inside the solar system, this leads to the
potential
\beq
\Phi_{\odot}(r) \; = \; - \,
{\displaystyle \frac{G \, M_{\odot}}{r}} \,
\left\{ 1 \, - \, {\displaystyle \frac{r^{2}}{2 \, L^{2}}} \right\}
\;\; ,
\eeq
and to the gravitational field
\beq
g(r) \; = \; - \, {\displaystyle \frac{G \, M_{\odot}}{r^{2}}}
\, - \, {\displaystyle \frac{G \, M_{\odot}}{2 \, L^{2}}} \;\; .
\eeq
Should the solar system be embedded inside the Bose condensate of
the field under scrutiny in this article, the various planets and
satellites that orbit around the Sun should undergo the additional
constant radial attraction
\beq
\delta g \; = \; 2.85 \times 10^{-18} \; {\rm cm \, s^{-2}} \;\;
\left\{ {\displaystyle \frac{10^{-2}}{\lambda}} \right\} \;
\left\{ {\displaystyle \frac{m}{1 \, {\rm eV}}} \right\}^{4} \;\; .
\eeq
This acceleration is so weak that it should not alter the motion
of the planets around the Sun. The relative increase of the solar
gravitational attraction is actually
\beq
{\displaystyle \frac{\delta g}{g}} \; = \;
1.93 \times 10^{-17} \;\;
\left\{ {\displaystyle \frac{r}{\rm 1 \, AU}} \right\}^{2} \;
\left\{ {\displaystyle \frac{10^{-2}}{\lambda}} \right\} \;
\left\{ {\displaystyle \frac{m}{1 \, {\rm eV}}} \right\}^{4} \;\; .
\eeq

\vskip 0.1cm 
\noindent
Detailed analysis of radio metric data from Pionner 10 and 11 indicate
the existence of an apparent anomalous acceleration acting on these
spacecrafts \cite{pioneer}. Quite exciting is the observation that this
anomalous acceleration $\delta g_{P}$ is constant and directed towards the
Sun. Both features are actually expected in the presence of a self--interacting
scalar field. However, the magnitude of the observed anomalous acceleration
$\delta g_{P} \sim 8.5 \times 10^{-8}$ cm s$^{-2}$ is ten orders of
magnitude larger than what is needed to explain the rotation curve
of DDO~154. Should the Pioneer acceleration be the consequence of a
scalar field enhanced solar gravity, it would indicate an exceedingly
large value for $m / \lambda^{1/4}$ of order 1.3 keV and a typical
condensate size $L \sim 9 \times 10^{-3}$ pc. We are therefore
lead to the conclusion that we cannot explain with the same value of
$m / \lambda^{1/4}$ the Pioneer anomalous acceleration and the 
rotation curves of dwarf spirals.

\section{Cosmological behavior}
\label{sec:cosmology}

\noindent We will now consider briefly the possible cosmological 
behavior of our scalar field. In Paper~II, we studied the cosmological
evolution of a homogeneous complex scalar field with a quadratic
and/or quartic potential. Here, we want to update this analysis for
the values of $m^4 / \lambda$ found in section~\ref{sec:rotation_curves}.
Generally speaking, focusing on the homogeneous quantities is the
first step in any comprehensive study of a given cosmological
scenario. In our case, we need to know whether the evolution of the
field background violates any cosmological bound before studying the
possible growth of spatial fluctuations -- hoping that they will
cluster and form galactic halos after the time of equality between
radiation density and field density.

\vskip 0.1cm
\noindent
We refer the reader to Paper~II for a detailed resolution of the
Klein--Gordon and Friedman equations in a Universe containing ordinary
radiation, baryons, a homogeneous complex field and a
cosmological constant relevant only today.  It is straightforward to
show that when the potential is dominated by the quartic term, the
energy density of the field smoothly decays as $a^{-4}$: so, in the
early Universe, the scalar field behaves as ``dark radiation''.
Later, when the quadratic term takes over, i.e., when
\begin{equation}
m^2 |\phi|^2 \sim \lambda \, |\phi|^4
\qquad \Rightarrow \qquad
V(\phi) \, \sim \, 2 \, \frac{m^4}{\lambda},
\end{equation}
the field starts to decay as $a^{-3}$, like dark matter: so, it can be
responsible for a ``matter-like'' dominated stage. During the whole
cosmological evolution, the kinetic energy of the field
is of the same order of magnitude as its potential energy. So, the
ratio $m^4 / \lambda$ immediately gives a rough estimate of the
total energy density of the field at the time of its transition,
denoted later as $\rho_{\phi}^{tr}$.  If this density is $\sim$
1~eV$^4$, we immediately notice that it is of the same order of
magnitude as the density at radiation--matter equality --
remember that $\rho_{\rm eq} \simeq 0.55$~eV$^4$ for the concordance
$\Lambda$CDM model.
So, in the early Universe, the density of our ``dark radiation'' (the
scalar field) had to be comparable to that of ``true radiation''
(photons and neutrinos). This brings some considerable tension with
the bound on the total radiation density that can be derived from BBN.

\vskip 0.1cm
\noindent
This cosmological toy--model and the problems associated with it were
first discussed by Peebles \cite{peebles} in the case of a real scalar
field, with essentially the same motivations as in the present
work. As a possible way out, Peebles proposed a small modification of
the scalar potential,
\begin{equation}
V(\phi) = m^2 \phi^2 + \lambda \phi^{q},
\end{equation}
where $q$ would be non--integer and slightly smaller than 4.
Indeed, by lowering the index $q$, one can decrease the fraction of
dark radiation in the early Universe, and in particular at BBN.
We will not follow this direction. Indeed,
the analysis of section~\ref{sec:rotation_curves} revealed a preferred
value of $m^4 / \lambda$ around 50--75~eV$^4$. This is significantly
larger than the observed value of $\rho_{\rm eq}$ and than the rough
estimate of Paper~I where we considered $m^4 / \lambda \simeq 1$~eV$^4$.
Our purpose in the rest of this section will be to check
whether this new value is compatible with the BBN bound.

\vskip 0.1cm
\noindent
In order to obtain a precise relation between the parameter
$m^4 / \lambda$ and the effective number of neutrinos at BBN -- which is
a convenient way to parameterize a cosmological density which
behaves like some extra relativistic degrees of freedom -- we need to
study numerically the detailed behavior of the field in a vicinity of
the transition between the
radiation--like and the matter--like regimes. For each value of
$m^4 / \lambda$, it is possible to follow $\rho_{\phi}$, and to
extrapolate
the branches in $\rho_{\phi} \propto a^{-4}$ and in
$\rho_{\phi} \propto a^{-3}$. We define $\rho_{\phi}^{\rm tr}$
as the energy density given by intersecting the two
asymptotes. The knowledge of this single number
is sufficient in order to relate exactly the constant value of
$\rho_{\phi} a^4$ in the early Universe to the constant value of
$\rho_{\phi} a^3$ measured today. A simple numerical simulation
gives
\begin{equation}
\rho_{\phi}^{\rm transition} = 2.4 \; m^4 / \lambda
\end{equation}
independently of any other field or cosmological parameters.
The simulation also provides a very good analytic approximation of the
field density at any time -- imposing that today, when
$a=a_0$, the field density is given by the fraction of the critical
density usually attributed to Cold Dark Matter,
$\rho_{\phi}= \Omega_{\rm cdm} \rho_c^0$:
\begin{equation}
\rho_{\phi} = \Omega_{\rm cdm} \rho_c^0
\left[ (a_0/a)^6 +
\left(\Omega_{\rm cdm} \rho_c^0  \, \frac{\lambda}{2.4 \, m^4}
\right)^{2/3}
(a_0/a)^8 \right]^{1/2} \;\; .
\end{equation}
The field density before the transition can be read directly from the
previous
equation. It can be conveniently parameterized in terms of an effective
neutrino number, defined as usual through
\begin{equation}
\Delta N_{\rm eff} = \frac{\rho_{\phi}}{\rho_{\nu}} \;\; ,
\end{equation}
where $\rho_{\nu}$ is the standard density of a single relativistic 
neutrino species. The final result is
\begin{equation}
\Delta N_{\rm eff} = \left(
\frac{\Omega_{\rm cdm}^4 \, \, \rho_c^{0 \,4} \, \, \lambda}
{\, \, \, \, \, 2.4 \, \, \, \, \, \, \rho_{\nu}^{0 \, 3} \, \, m^4}
\right)^{1/3} = 7.5 \left( \frac{\Omega_{\rm cdm} h^2}{0.13} 
\right)^{4/3}
\left( \frac{\lambda^{1/4} eV}{m} \right)^{4/3} \;\; .
\end{equation}
Sticking to $m^4 / \lambda = 50~({\rm eV})^4$, and using the currently
preferred values $h = 0.68$ and $\Omega_{\rm cdm} = 0.3$, one finds
$\Delta N_{\rm eff} = 2$, which is above the usual BBN bound
$|\Delta N_{\rm eff}| < 1$ \cite{bbn}. However, it is still possible to
find some values of ($\Delta N_{\rm eff}$, $h$, $\Omega_{\rm cdm}$)
satisfying the above relation, and allowed at the 1--$\sigma$ level by
current CMB experiments and BBN predictions \cite{archeops} --
for instance, (1.0, 0.63,0.20).
In any case, in a very near future, the new CMB observations will
set some stringent limits on these three parameters: it will then
be easy to state about the validity of our alternative to the usual 
cosmological scenario.

\section{Conclusions}
\label{sec:conclusion}

\noindent
We have shown that a self--coupled
charged scalar field provides an excellent fit to the
rotation curve of the dwarf spiral DDO 154. That galaxy
is the prototypical example of a system known to be completely
dominated by dark matter. The effect of the quartic coupling
$\lambda$ results into an effective modified
gravitation inside the Bose condensate where the Poisson equation
becomes strongly non--linear. The problem -- complicated by the
non--sphericity of the baryon distribution -- has been solved
exactly as explained in section~\ref{sec:method}. The agreement
with the observations of the
circular speed of DDO 154 is impressive. Notice that neutralino
dark matter does not pass this test because of the central cusp
that it would develop. We conclude that the charged scalar field
considered in this analysis provides an exciting alternative to
the galactic dark matter -- at least inside dwarf systems.
A typical value of $m^{4} / \lambda \sim$ 50 -- 75 eV$^{4}$
obtains.

\vskip 0.1cm
\noindent
The scalar field behaves cosmologically as a dark radiation component
as long as the quartic contribution of the potential
$V \left( \phi \right)$ dominates over its quadratic counterpart.
The situation gets reversed when the field energy density is
$\sim 2.4 \; m^{4} / \lambda$ and a matter--like behavior subsequently
ensues. The larger the crucial parameter $m^{4} / \lambda$, the sooner
the transition between dark radiation and dark matter--like behaviors
and consequently the smaller the contribution of the scalar field to
the overall radiation density at early times -- for a fixed scalar field
mass density today. A large value of $m^{4} / \lambda$ translates into
a small number of effective neutrino families during BBN and we
have shown that our model marginally satisfies the requirement that
$\Delta N_{\rm eff}$ should not exceed 1.

\vskip 0.1cm
\noindent
Actually the model is strongly constrained on the one hand
side by the size $L$ of the Bose condensates -- and therefore of the
corresponding galactic halos -- and on the other hand side by the
contribution $\Delta N_{\rm eff}$ to the radiation density at BBN.
Both $L$ and $\Delta N_{\rm eff}$ decrease with $m^{4} / \lambda$
and a value for the latter of $\sim$ 50 -- 75 eV$^{4}$ -- which provides
excellent agreement with DDO 154 -- is marginally consistent with BBN.
Large halos cannot consequently been pictured in terms of a single Bose
condensate and the simple scheme presented here has to be modified. 
%
A possible solution -- yet not very natural -- is to replace the
quartic field self--interaction by a $\phi^{q}$ term as suggested
by \cite{peebles}. This would alleviate the BBN constraint.

\vskip 0.1cm
\noindent
Another option worth being explored is to imagine that massive
and extended halos are formed of several bosonic clumps. The
coherent configuration that has been investigated here may be
understood as the ground state of some gigantic bosonic atom.
It is therefore conceivable that the scalar field may also form
several such condensates that would be organized inside a huge
bosonic molecule with a spatial extension much in excess of $L$.
The electron cloud around the proton does not extend further
than $\sim 10^{-10}$ m inside the hydrogen atom and yet
electrons are delocalized over meter size distances inside metals.

\vskip 0.1cm
\noindent
If so, the dark matter would be made of small bosonic clumps.
Should the solar system lie within such a system, the
motion of its planets would provide in that case a lower bound
on $L$ since the smaller is the latter, the stronger is the
effective modification to Newton's law of gravitation. We have
actually shown in section~\ref{sec:solar_system} that
the scalar field concentrates in the solar potential well
and strengthens it to generate an additional gravitational
attraction that is radial and constant. As a matter of fact,
the radio data from the Pioneer probes are consistent with
such an anomalous acceleration that seems to be constant
and directed towards the Sun. Assuming that it results from
the self--interacting scalar field which we have investigated
in this work, the observed magnitude
$\delta g_{P} \sim 8.5 \times 10^{-8}$ cm s$^{-2}$ would imply
a value for $m / \lambda^{1/4}$ of order 1.3 keV and a typical
condensate size $L \sim$ 0.01 pc. The merging of many of these
small bosonic clumps into a larger structure like a galactic halo
is an open question.


\section*{Acknowledgments}
\noindent
We would like to thank D.~Maurin and R.~Taillet for useful 
discussions.



\appendix
\section{}
\label{sec:appendixI}

\vskip 0.1cm
\noindent 
The purpose of this section is to derive the gravitational potential
generated by a pure scalar field condensate.
This calculation is complementary to the one at the end
of section~\ref{sec:basic_equations}, based on the polytropic
formalism. In absence of a baryonic density $\rho_b$,
eq.~(\ref{poisson1}) reads
\beq
\Delta \left(\Phi-\Phi_0\right) + 8 \pi G\frac{m^4}{\lambda} 
\left(\Phi-\Phi_0\right) {\cal H}\left(\Phi_0-\Phi\right) = 0 \, .
\eeq
\noindent Since there is no source, let us suppose that the 
gravitational 
potential has a spherical symmetry. One can do a change of variable 
$r=z\sqrt{\frac{\lambda}{8 \pi G m^4}}$ so that the equation becomes 
simply 
\beq
\frac{d^2}{dz^2}\left\{z \, (\Phi-\Phi_0)\right\} + z \, 
(\Phi-\Phi_0) \, \,
{\cal H}\! \left(\Phi_0 \! - \! \Phi\right) = 0 \, .
\eeq
\noindent The only solution of this equation 
which is continuous and derivable everywhere, 
and goes to zero at infinity is
\beq
\Phi(z) =\left\{\begin{array}{ll}
\displaystyle \Phi_0 + (\Phi(0)-\Phi_0) \frac{\sin z}{z} & {\rm if} 
\;\; z < \pi \qquad {\rm (see~Paper~II)}\\
\\
\displaystyle \frac{\pi \Phi_0}{z} & {\rm otherwise.}
\end{array}\right.
\label{eq:A3}
\eeq
\noindent So, the maximum extension of the scalar field
halo is $\displaystyle 
r_{\rm max}=\pi \sqrt{\frac{\lambda}{8 \pi G m^4}}$. The density of 
the bosonic halo is
\beq
\rho_\phi = -\frac{2 m^4}{\lambda} \left(\Phi-\Phi_0\right) 
{\cal H}\left(z - \pi\right)
\eeq
\noindent and the total mass is
\beq
M = 4 \pi \left(\frac{\lambda}{8 \pi G m^4}\right)^{3/2} 
\int_0^{+\infty} \rho_\phi(z') {z'}^2 dz'= \frac{\pi}{G} 
\sqrt{\frac{\lambda}{8\pi G m^4}} \left(\Phi_0-\Phi(0)\right) \, .
\eeq
\noindent On the other hand, the Gauss theorem applied to
the sphere of radius $r = r_{\rm max}$ gives
\beq
\Phi_0=-\frac{GM}{r_{\rm max}}=- \frac{GM}{\pi}\sqrt{\frac{8\pi G 
m^4}{\lambda}}
\eeq
\noindent so that $\Phi(0)=2 \Phi_0$. The solution (\ref{eq:A3}) 
can be rewritten as
\beq
\Phi(z) = \left\{\begin{array}{ll}
\displaystyle \Phi_0 \left(1+ \frac{\sin z}{z} \right) & {\rm if} 
\;\; z < \pi\\
\\
\displaystyle \frac{\pi \Phi_0}{z} & {\rm otherwise.}
\end{array}\right.
\eeq

\section{}
\label{sec:appendixII}

\vskip 0.1cm
\noindent The purpose of this Appendix is to provide a prescription
for the definition of $\Phi^{(0)}$ -- the starting function in the
recursive method. We tested this prescription on
various examples, and found that $\Phi^{(0)}$ is always a fairly good
approximation of the exact solution, allowing for quick convergence.

\vskip 0.1cm
\noindent
The idea is to enforce the boundary surface on which the
field density
$(m^4 / \lambda) 
(\Phi-\Phi_0) \, \, {\cal H}\! \left(\Phi_0 \! - \! \Phi\right)$
vanishes to be a perfect sphere. 
Of course, this has to be wrong when the baryonic density
is non--spherical. So, if we impose a constant boundary radius,
we need to relax the fact that the value of 
$\Phi$ should be constant all over the 
boundary. In other terms, if $r$ is the radial coordinate, we replace
${\cal H} \! \left(\Phi_0 \! - \! \Phi\right)$ by
${\cal H} \! \left(r_0 \! - \! r\right)$, where $r_0$ is an
arbitrary boundary radius.

\vskip 0.1cm
\noindent
If we define a dimensionless radial coordinate
$z=r\sqrt{\frac{8 \pi G m^4}{\lambda}}$, eq.~(\ref{poisson1}) becomes
\beq
\label{eq:modifequation}
\Delta \Phi = S(z,\theta) + \left(\Phi_0-\Phi\right) 
{\cal H}\left(z_0-z\right).
\eeq
\noindent where $S = 4 \pi G \rho$. After a
Legendre transformation, we obtain the following set of 
differential equations:
\beq
\frac{1}{z^2}\frac{d}{dz}\left( z^2 \frac{d\Phi_l}{dz}\right) - 
\frac{l(l+1)}{z^2} \left(\Phi_l -\Phi_0 \delta_{l,0} \right) 
{\cal H}(z_0-z) = S_l(z)
\eeq
\noindent where $\delta_{l,0}$ is the Kronecker symbol.
Each of these equations can be solved separately on the two intervals 
$0 \leq z \leq z_0$ and $z_0 \leq z \leq \infty$, using Green 
functions.
In terms of the spherical Bessel functions $j_l(z)$ and $n_l(z)$,
the solution of eq.~(\ref{eq:modifequation}) for $z \le z_0$ is
\beq
\Phi_l(z)=\Phi_0 \delta_{l0} + n_l(z) \int^z_0 u^2 j_l(u) S_l(u) du 
+  j_l(z) \int_{z}^{z_{0}} u^2 n_l(u) S_l(u) du + C_l^1 j_l(z)
\label{eq:B4}
\eeq
\noindent where $C_l^1$ is a free constant. In the same way, 
for $z \ge z_0$, the solution is
\beq
\Phi_l(z)=-\frac{1}{2l+1} z^{-(l+1)} \int_{z_0}^z u^{l+2} S_l(u) du 
-\frac{1}{2l+1} z^l \int_z^{+\infty} u^{1-l} S_l(u) du 
-C_l^2\frac{1}{2l+1} z^{-(l+1)}
\label{eq:B5}
\eeq
\noindent where $C_l^2$ is another free constant. We impose
that each Legendre coefficient of the gravitational potential is derivable and 
continuous on the boundary $z=z_0$:
\beq\left\{
\begin{array}{lll}
\Phi_l(z_0^-) & = & \Phi_l(z_0^+) \, ,\\
\\
\displaystyle \frac{d}{dz}\Phi_l(z_0^-) & = &\displaystyle  
\frac{d}{dz}\Phi_l(z_0^+) \, .
\end{array}
\right.
\eeq
\noindent This defines a unique value for each constant of integration:
\beq
\begin{array}{lll}
C_l^1 & = & \displaystyle \frac{(I^2_l-I^1_l 
- \Phi_0 \, \delta_{l,0})(l+1)+z_0 
(J^2_l-J^1_l)}{(l+1)j_l(z_0)+z_0 j'_l(z_0)} \, ,\\
\\
C_l^2 & = & \displaystyle \frac{-z_0^{l+2} (2l+1) j_l(z_0) 
(J^2_l-J^1_l)+z_0^{l+2} (2l+1) j'_l(z_0) 
(I^2_l-I^1_l - \Phi_0 \, \delta_{l,0})}{(l+1)j_l(z_0)+z_0 j'_l(z_0)} \, ,\\
\end{array}
\eeq
\noindent where
\beq
\begin{array}{lll}
I_l^1 &=& \displaystyle n_l(z_0) \int_0^{z_0} 
u^2 j_l(u) S_l(u) du \, ,\\
\\
I_l^2&=& \displaystyle -\frac{1}{2l+1} z_0^l \int_{z_0}^{+\infty} 
u^{1-l} S_l(u) du \, ,\\
\\
J_l^1 &=& \displaystyle n'_l(z_0) \int_0^{z_0} u^2 j_l(u) S_l(u) du \, ,\\
\\
J_l^2&=& \displaystyle -\frac{l}{2l+1} z_0^{l-1} \int_{z_0}^{+\infty} 
u^{1-l} S_l(u) du \, .\\
\end{array}
\eeq
\noindent One can then reconstruct the gravitational potential 
$\Phi^{(0)}$
from
\beq
\Phi^{(0)}(z,\theta)=\sum_{l=0}^{+\infty}P_l(\cos \theta) \Phi_l(z) \, .
\label{eq:B9}
\eeq
\noindent 
So far, the approximate solution $\Phi^{(0)}$ constructed in 
this way depends on two arbitrary numbers: first, $z_0$, and second,
$\Phi_0$, which appears explicitly in the definition of $C_0^1$ and
$C_0^2$. However, $z_0$ and $\Phi_0$ have to be related in some way.
Indeed, if $\Phi^{(0)}$ was an exact solution, $\Phi_0$ would be equal
by definition to $\Phi^{(0)}(z_0, \theta)$ for any $\theta$.
In our approximation scheme, $\Phi^{(0)}(z_0, \theta)$ is not 
independent of $\theta$, but we can choose
a particular direction $\theta_0$, and impose that
$\Phi_0 = \Phi^{(0)}(z_0, \theta_0)$. Inserting this identity in 
eq.~(\ref{eq:B9}), and using eq.~(\ref{eq:B4}), one obtains the 
relation
\beq
\Phi_0 = \Phi^{(0)}(z_0, \theta_0) = \frac{z_0}{\tan z_0} 
\sum_{l=0}^{+\infty} P_l(\cos \theta_0) \left\{ 
n_l(z_0) \int_0^{z_0} u^2 j_l(u) S_l(u) du + K_l \, j_l(z_0) \right\}
\label{eq:phionboundary}
\eeq
\noindent where
\beq
K_l = \frac{(I^2_l-I^1_l)(l+1)+z_0 
(J^2_l-J^1_l)}{(l+1)j_l(z_0)+z_0 j'_l(z_0)} \, .
\eeq
\noindent In summary,
the first step of our recursive method is performed in the following 
order:
\begin{enumerate}
\item
we choose a value $z_0$ (or $r_0$) and an arbitrary direction $\theta_0$ 
(that will be kept for all the following iterations).
\item
we solve eq.~(\ref{eq:phionboundary}) in order to find
$\Phi_0$.
\item
we compute $\Phi^{(0)}(z, \theta)$ using 
eqs.~(\ref{eq:B4}), (\ref{eq:B5}), (\ref{eq:B9}).
\end{enumerate}
The next iterations are performed in the much simpler way 
described 
in section~\ref{sec:method}.

\end{document}